\documentclass[aps,
prx,
longbibliography,notitlepage,reprint,superscriptaddress]{revtex4-2}
\usepackage[colorlinks,linkcolor=blue,citecolor=blue,urlcolor=red]{hyperref}
\usepackage{soul}
\usepackage{changes}

\usepackage{float}
\usepackage{CJK}
\usepackage{graphicx}%,float}
\usepackage{amsmath,amssymb,amsfonts}
\usepackage{siunitx}
\usepackage{soul}
\usepackage{chngcntr}

\usepackage{amsmath,amssymb,amsfonts} % standard AMS packages
\usepackage{bm}	% bold symbols in math mode \bm{...}
\usepackage{dsfont}	% proper mathbb format
\usepackage{mathrsfs} % use \mathscr{} for script letters in math
\usepackage{mathtools} % for proper typesetting of := and =:
\newcommand{\bd}[1]{\mathbf{#1}}
\newcommand{\SiN}{$\mathrm{Si_3N_4}$}
\newcommand{\GOE}{Georg-August-Universit\"{a}t G\"{o}ttingen, D-37077 G\"{o}ttingen, Germany}
\newcommand{\EPFL}{Swiss Federal Institute of Technology Lausanne (EPFL), CH-1015 Lausanne, Switzerland}
\newcommand{\MPIBPC}{Max Planck Institute for Multidisciplinary Sciences, D-37077 G\"{o}ttingen, Germany}
\newcommand{\TAI}{School of Electrical Engineering, Tel Aviv University, 69978, Tel Aviv, Israel}
\newcommand{\EPFLQ}{Center for Quantum Science and Engineering, EPFL, CH-1015 Lausanne, Switzerland}
\begin{document}

\author{Guanhao Huang}
\email{guanhao.huang@epfl.ch}
\affiliation{\EPFL}
\affiliation{\EPFLQ}
\author{Nils J. Engelsen}
\affiliation{\EPFL}
\affiliation{\EPFLQ}
\author{Ofer Kfir}
\affiliation{\TAI}
\author{Claus Ropers}
\affiliation{\MPIBPC}
\affiliation{\GOE}
\author{Tobias J. Kippenberg}
\affiliation{\EPFL}
\affiliation{\EPFLQ}

\title{Quantum state heralding using photonic integrated circuits with free electrons}

\begin{abstract}
Recently, integrated photonic circuits have brought new capabilities to electron microscopy and been used to demonstrate efficient electron phase modulation and electron-photon correlations. Here, we quantitatively analyze the feasibility of high fidelity and high purity quantum state heralding using a free electron and a photonic integrated circuit with parametric coupling, and propose schemes to shape useful electron and photonic states in different application scenarios. Adopting a dissipative quantum electrodynamics treatment, we formulate a framework for the coupling of free electrons to waveguide spatial-temporal modes. To avoid multimode-coupling induced state decoherence, we show that with proper waveguide design, the interaction can be reduced to a single-mode coupling to a quasi-TM\textsubscript{00} mode. In the single-mode coupling limit, we go beyond the conventional state ladder treatment, and show that the electron-photon energy correlations within the ladder subspace can still lead to a fundamental purity and fidelity limit on complex optical and electron state preparations through heralding schemes. We propose applications that use this underlying correlation to their advantage, but also show that the imposed limitations for general applications can be overcome by using photonic integrated circuits with an experimentally feasible interaction length, showing its promise as a platform for free-electron quantum optics.
	
\end{abstract}
\maketitle

Quantum coherent coupling between distinct physical systems harnesses the advantages and strengths of the different systems in order to better explore new phenomena and potentially develop novel quantum technologies~\cite{Kimble2008,Wallquist_2009}. 
Photonic links~\cite{QuICs} are most commonly used to connect different systems due to the potential for long-range transmission through optical fibers and robustness to decoherence from thermal environments, and have been realized in systems ranging from superconducting qubits~\cite{Mirhosseini2020,Andrews2014}, solid state spins~\cite{Sipahigil2016,Wan2020}, ultra coherent mechanics~\cite{Gregory20,Riedinger2018}, and atomic systems~\cite{ionIC,Brand13,itay14,Tiecke2014,Brekenfeld2020}, where each offers unique features and advantages to be utilized in a hybrid quantum system.
One key aspect of all these systems is the ability to enact high-fidelity quantum control of the interaction with well-defined optical modes. 

In the field of electron microscopy, interactions between free electrons and photons have been widely explored in both stimulated ~\cite{Barwick2009, GarciadeAbajo2010, Yurtsever2012, Feist2015, Piazza2015, Wang2020_cavity, Kurman2021, Liebtrau2021} and spontaneous processes \cite{Kociak2017,Polman2019, GarciadeAbajo2021_review, GarciadeAbajo2010} enhanced by phase-matched interactions and optical resonances
\cite{Kfir2020, Muller2021, Auad2022,Pomarico2018,Henke2021,Shiloh2021}.
%On the single photon level, measurements of quantum statistics that have been recently introduced to electron microscopy~\cite{Meuret2015, Bourrellier2016}. 
There have also been many proposals which explore the unique quantum properties of electron-photon states~\cite{Priebe2017, Kfir2019, DiGiulio2019, BenHayun2021, Yalunin2021, Dahan22}. However, it is still an open question whether high-fidelity quantum control of this hybrid quantum system can be realized.

High-fidelity quantum control requires high coupling strength between free electrons and optical vacuum fields, and low dissipation to keep decoherence at a minimum. The interaction mechanisms and their coupling strengths differ substantially between different physical platforms, which can be categorized into metallic~\cite{Feist2015} and dielectric structures~\cite{Kfir2020, Wang2020_cavity, Henke2021,Kozak2017_acceleration, Dahan2020}.
For nanophotonic particles, the short attosecond-long interaction time promotes the use of dissipative materials, such as plasmonic structures~\cite{Gramotnev2010}. The collective electronic response amplifies the interaction, while at the same time bringing retardation and dissipation, which is not ideal for quantum-coherent manipulation of electrons with optical states. On the other hand, transparent dielectrics, for which the coupling is enhanced by an extended interaction length, offer a paradigm shift in free-electron quantum optics due to their low optical dissipation and practically instant electronic response. Instead of enhancing the interaction by lossy media, optical modes supported by dielectric structures interact with the free electron by a geometric effect through the relativistic field retardation~\cite{Abajo1998}, which results in a purely parametric interaction ideally suited for high-fidelity quantum control.

Photonic integrated circuits have only entered the picture very recently~\cite{Henke2021,feist22}, and have several advantages for free-electron quantum optical experiments. Firstly, integrated photonics enables exquisite control of the optical properties of waveguides~\cite{Pfeiffer:17,Moille22}. 
The nearly lossless guided modes~\cite{Liu2021} and high-efficiency output fiber coupling~\cite{Almeida03} facilitate coupling to both on-chip ~\cite{Sipahigil2016,Wan2020,Gregory20,Riedinger2018}  and fiber-coupled quantum systems ~\cite{Brand13,itay14,Tiecke2014,Brekenfeld2020,Mirhosseini2020}. Additional capabilities are provided by well-established on-chip optical elements such as tunable beam splitters and phase shifters~\cite{Bogaerts2020}, spectral filters~\cite{Li2021} and photon counters~\cite{Najafi2015}, which offer high-fidelity optical state manipulation and characterization~\cite{Politi2008}. With the versatile on-chip structures and demonstrated efficient electron phase modulation~\cite{Henke2021} and electron-photon correlation~\cite{feist22}, we propose heralding schemes to shape useful electron and optical states in various application scenarios with photonic integrated circuits.

\begin{figure}[t]
 	\includegraphics[width = 0.49\textwidth]{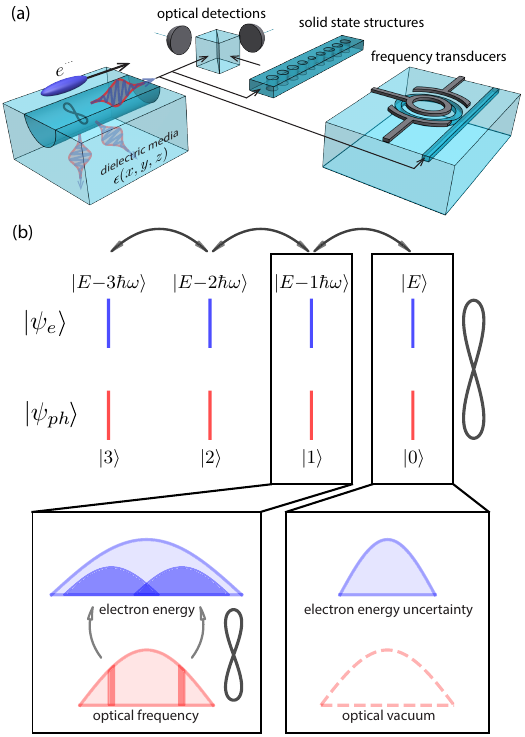}	\caption{\textbf{(a)} Illustration of the electron-photon inelastic scattering process mediated by a dielectric waveguide. In an electron microscope, when the high-energy electron passes by a dielectric waveguide structure with a given dielectric permittivity distribution $\epsilon(x,y,z)$, the material dielectric dipoles exert a backaction field (force) on the electron, resulting in correlated electron energy loss and optical emission in both the guided waveguide modes and non-guided bulk modes. High-ideality coupling to a low-loss waveguide mode is required for high fidelity state preparation and interaction with other quantum systems through optical links. \textbf{(b)} Synthetic electron-photon state ladder of the pair state generation through $\hat S_{\mathrm{e\text{-}ph}}$. Within each ladder state, there is an underlying subspace that still maintains correlation between electron energy and photon frequencies. This correlation can lead to new types of applications, but generally leads to degradation of fidelity and purity of the interaction. }
	\label{fig:illus}
\end{figure}

To transfer the aforementioned advantages to the scenario of generating high-quality quantum states through electron-photon interaction, high-ideality coupling to a single well-defined optical mode~\cite{Bendana2011} is required. However, due to the complex waveguide structures, parasitic couplings to auxiliary spatial modes cause decoherence of the system, see Fig.\ref{fig:illus}(a). We quantitatively investigate this limitation in a realistic experimental scenario, and show that with a single-mode waveguide, larger gap distance, and long interaction length, near-unity coupling ideality and strong coupling can be achieved to the waveguide quasi-TM\textsubscript{00} spatial-temporal mode. 

Further, we show that even in the limit of single-mode interaction, there is still a state subspace correlation that imposes fundamental limit to the state fidelity and purity. To address the electron-photon interaction in the conventional quantum optics description, a synthetic ladder state space~\cite{Priebe2017, Kfir2019, DiGiulio2019, BenHayun2021, Yalunin2021, Dahan22} is usually used, shown in Fig.\ref{fig:illus}(b). This treatment greatly eases the analysis of the interaction between two systems that are actually continuum systems. However, within the subspace of a ladder level, energy conservation enforces strong correlation between the electron energy loss and the frequency of the photon created. When one neglects the underlying correlation, information loss occurs. This process can be characterized by the state purity, that captures both the distance to a pure quantum state, and the degree of electron-photon entanglement through R{\'e}nyi-2 entropy~\cite{renyi1961measures}. Here, we propose applications that exploit this underlying correlation to their advantage, e.g. imprinting electron wavefunctions onto optical states, and later examine the state fidelity and purity in quantum state heralding schemes. We find that electrons in particle-like states with high purity are required to generate pure heralded states, and the purity limits are greatly reduced with experimentally feasible interaction length using photonic integrated circuits.

The manuscript is organized as follows: Section I establishes the theoretical formalism for describing electron-photon spontaneous scattering processes with dielectric media, different parameter regimes, and the underlying state correlations within the energy ladders. Section II studies the interaction in a photonic integrated circuit structure, defines the spatial-temporal modes and provides guidelines to achieve single-mode coupling. Section III investigates optical state heralding in wave-like and particle-like regimes, shows corresponding applications and quantitative analysis of the correlation-induced state heralding fidelity and purity limit. Section IV investigates electron state heralding schemes, optical mode matching and down-conversion schemes, and the state purity limit. Section V summaries the manuscript, discuss the theoretical limitations of our analysis and the experimental constraints.

\section{Electron-photon interactions with dielectric media}

\begin{figure*}[t]
 	\includegraphics[width = 1\textwidth]{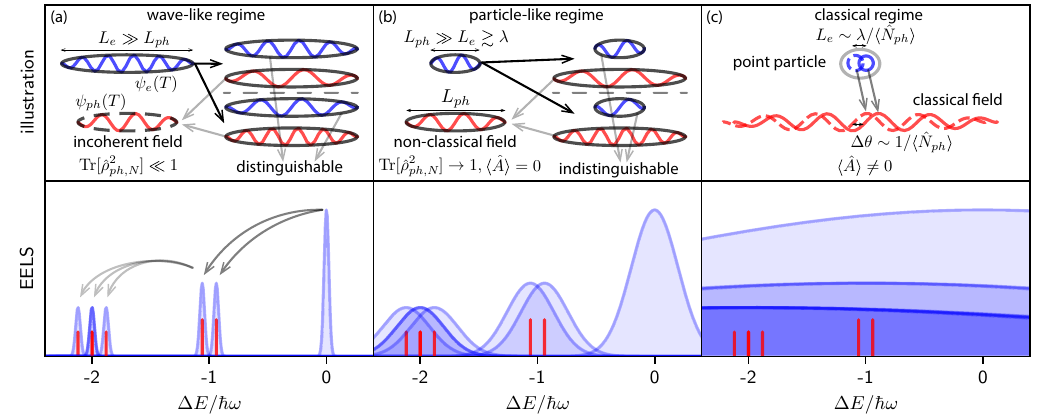}	\caption{Illustration of different parameter regimes of electron-photon interactions. The electron and photon spatial waveforms as a function of $T=t-z/v_e$, where $z$ is the longitudinal coordinate, are shown in the upper panels, and the corresponding energy (frequency) domain picture is shown in the lower panels. The frequency bandwidth of the generated photon is determined by the phase-matching mechanism, and two of the frequency components are shown in red. \textbf{(a)} Wave-like regime where distinguishable electron states are generated from the emission of photons with the corresponding frequencies. This regime is accompanied by mixed optical ladder states $\rho_{ph,N}$, and incoherent photon emission. \textbf{(c)} Particle-like regime where the photon emission at different frequencies generates indistinguishable electron states. This regime has pure optical ladder states $\rho_{ph,N}$, where the synthetic electron-photon state ladder is a valid approximation. \textbf{(c)} Classical regime where different photon-sidebands of the electron overlap well, and classical optical field emission with non-zero $\langle \hat A\rangle$ is achieved. }
	\label{fig:theory}
\end{figure*}

The interaction between free electrons and optical modes at a dielectric surface can be understood in a microscopic picture as follows: When an electron passes near the surface of a dielectric structure, the electric field of the flying electron polarizes the dipoles in the structure  (see Fig.\ref{fig:illus}(a)). As a result, these dipoles generate oscillating electromagnetic fields that cause backaction Coulomb forces on the electron which change the electron energy. In the conventional quantum optical modal decomposition picture commonly used in the cavity quantum electrodynamics (QED) community, this can be interpreted as the interaction between the free electrons and the optical vacuum fields of the modes supported by the dielectric structure~\cite{milonni1994quantum}.

Here, we formulate the problem as the interaction between propagating free electrons and one single interaction-specific optical spatial mode $\hat{\bd{A}}(\bd{r},\omega)$ (see Appendix \ref{appendix: theory} for the QED details and the field profile) at frequencies $\omega$ in the continuum, instead of pre-determined discrete optical modes of the dielectric structure (see Appendix \ref{append:MD} for its correspondence to modal decomposition), with the scattering matrix~\cite{Bendana2011,Kfir2019} in the interaction picture
\begin{equation}
	\hat S_{\mathrm{e\text{-}ph}} = e^{i\hat \chi}\exp\left[\int d\omega g_\omega\hat b_\omega^\dag \hat a_\omega-h.c.\right],\label{eq:scatter}
\end{equation}
where the phase operator $\hat\chi$ acts only on the electron degrees of freedom (ignored in the remaining discussion), and is associated with the Aharonov–Bohm effect of the vector potential~\cite{Barwick_2008}. Continuum photon ladder operators $\hat a_\omega$ and electron operators $\hat b_\omega$ characterize the energy exchanges between the electron and the optical field at a given optical frequency $\omega$ in an energy-conserving manner. The interaction with the vacuum optical fields results in transitions into lower electron energy states with energy differences of $\hbar\omega$. We define the electron-photon coupling strength at a given photon energy $\hbar\omega$ in terms of the vacuum coupling strength $g_\omega$ as $\Gamma(\omega) = |g_\omega|^2$~\cite{GarciadeAbajo2010}. The phase matching condition gives the vacuum coupling strength a finite bandwidth. In the limit where $\Gamma(\omega)\ll 1$, $\Gamma(\omega)$ is equivalent to the electron energy loss probability per unit optical frequency of dielectric media measured in electron energy loss spectroscopy (EELS), and can be derived classically in a simplified picture (See Appendix~\ref{append:class}). To simplify the discussion here, we also assume a point-like transverse distribution for the electron (see Appendix \ref{appendix: theory} for the discussion on the transverse effect), and the vacuum coupling strength $g_\omega$ is derived at a corresponding transverse position $\bd{R}_0$.

The interaction with the optical continuum, as opposed to the conventional discrete energy-ladder levels illustrated in Fig.\ref{fig:illus}(b), results in a continuum electron-photon pair state
\begin{gather}
    |\psi_e,\psi_{ph}\rangle = \exp\left(-\frac{\int d\omega|g_\omega|^2}{2}\right)\nonumber\\
    \times\left(\sum_n\frac{\left(-\int d\omega g_\omega^* \hat b_\omega \hat a_\omega^\dag\right)^n}{n!}\int dE\psi(E)|E\rangle|0\rangle\right),
\end{gather}
where $\psi(E)$ is the electron wavefunction in the energy domain. We show in Fig.~\ref{fig:theory} that depending on the size of the electron wave-packet, the electron-photon interaction can be categorized into three regimes. The classical regime has been explored, and is accessible through laser modulation schemes~\cite{Sergey21,Priebe2017,Morimoto2018,Norbert19,Yalunin2021,Kfir2021}. Some electron-microscopes equipped with a monochromator fall into the wave-like regime\cite{Krivanek19,Auad2022,Muller2021}, while others with longer interaction length~\cite{feist22} are in an intermediate wave-particle-like regime. The simplified electron-photon ladder picture is only partially valid in both cases. 

In the wave-particle-like regime, the ladder state
$ |\psi_e,\psi_{ph}\rangle_N \propto\int dE\psi(E){\left(\int d\omega g_\omega^* \hat b_\omega \hat a_\omega^\dag\right)^n}|E\rangle|0\rangle$ maintains a correlation between electron energies $E$ and photon frequencies $\omega$. To go back to the simplified ladder picture, one traces out e.g. the continuum electron states within each ladder as $\hat\rho_{ph,N}=\mathrm{Tr}_E\left[|\psi_e,\psi_{ph}\rangle_N\langle\psi_e,\psi_{ph}|_N\right]$, which results in a degradation of the optical state purity $\mathcal{P}=\mathrm{Tr}\left[\hat\rho_{ph,N}^2\right]$. To reduce the degree of correlation and reach the particle-like regime, a narrower phase-matching bandwidth relative to the electron energy uncertainty is generally required.

We show in the latter half of this manuscript that in the case of photonic integrated circuits, the prolonged interaction length can help reduce the phase-matching bandwidth and lower the energy correlation for a single waveguide mode, pushing the system parameters well into the particle-like regime. However, the complex dielectric environment generally results in multimode electron-photon interactions, e.g., through parasitic coupling to other optical mode families and other non-guided spatial modes supported by open-ended dielectric substrates. The effective phase-matching bandwidth of the multimode coupling is generally large, and the corresponding electron-photon correlation can not be suppressed by a longer interaction length. Therefore, we first quantitatively analyze how to effectively constrain the interaction to the single-mode case.

It is generally hard to design and fabricate waveguide structures that achieves $100\%$ spatial overlap between a waveguide mode and the electron optical emission over the full optical frequency range. Therefore, instead of mode-matching, our strategy to achieve single-mode interaction is to exploit a combination of effects which are results of the phase-matching mechanism. 

To quantitatively account for the infinite number of interacting spatial optical modes, it is generally impractical to use the conventional modal decomposition method~\cite{Feist2015}. Instead, as is mentioned before, we combine all the possible coupling contributions from different modes into one single interaction-specific spatial mode, following a three-dimensional QED treatment~\cite{3dquant}.
This formalism, derived using the fluctuation-dissipation theorem, was previously used when analyzing electron energy loss probabilities with dissipative materials~\cite{GarciadeAbajo2010}
% Here, we use it for dispersive dielectric materials, i.e. the studied wavelength range is far from material absorption bands, which simplifies the simulation by not requiring knowledge of the imaginary part of the dielectric permittivity.
that exhibit a delayed material response, which is the dominant contribution to the main electron energy loss channels. The dielectric materials we study here are transparent in the optical frequency bands of interest. In this sense, we can set $\mathrm{Im}\{\epsilon(\bd{r},\omega)\}\rightarrow 0$, which corresponds to an instantaneous dielectric dipole response and further simplifies the analysis. For materials with sufficiently low absorption, which are used for integrated waveguides designed to guide optical fields, the interaction is purely contributed from the relativistic field retardation effect~\cite{Abajo1998} and prohibits energy and momentum transfer to the material, avoiding loss of coherence. It is in this sense that the whole process of an electron interacting with dielectric waveguide is \emph{parametric} in nature.

\section{Coupling ideality}
\label{sec:ideality}

\begin{figure*}[t]
 	\centering
    \includegraphics[width=1\textwidth]{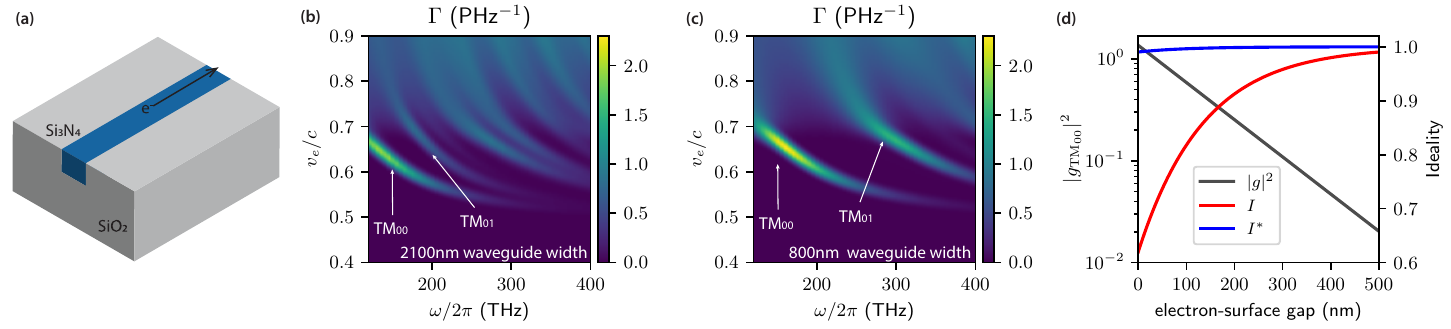}
    \caption{ \textbf{(a)} Illustration of the waveguide structure studied, consisting of a $\text{Si}_\text{3}\text{N}_\text{4}$ waveguide embedded in a silica substrate. The free electron passes by the top surface of the waveguide and generates correlated optical emission. \textbf{(b-c)} Electron-photon coupling strength $\Gamma(\omega)$ spectrum for different waveguide geometries and electron positioning. The coupling spectrum is plotted as a function of both electron velocity $v_e$ and optical frequency $\omega$. The waveguides have a thickness of \SI{650}{nm}, and widths of \textbf{(b)} \SI{2.1}{\micro m} and \textbf{(c)} \SI{800}{nm}. Coupling to different waveguide mode families appears as multiple coupling bands, and their phase-matching bandwidth is kept constant for better visualization. \textbf{(d)} Total coupling strength of TM\textsubscript{00} mode vs. non-conditional and conditional coupling idealities ($I$ and $I^*$ respectively), as a function of gap distance between the electron beam and the waveguide surface, with \SI{800}{nm} waveguide width, \SI{100}{\micro\meter} interaction length and $v_e/c=0.65$ electron velocity. 
    }
    \label{fig:fig2}
\end{figure*}
% \section{Quantitative analysis of electron-photon interaction}
In this section, we show how to achieve ideal single-mode electron-photon coupling with photonic integrated circuits. 
As an example, here we quantitatively investigate the electron-photon coupling mediated by an integrated \SiN waveguide embedded in a silica substrate without top cladding (the bottom silicon substrate is not considered), shown in Fig.~\ref{fig:fig2}(a). This type of structure has been used in recent investigations of both stimulated phase-matched interactions~\cite{Henke2021} and spontaneous inelastic scattering~\cite{feist22} between free electrons and the evanescent field of a photonic-chip-based optical microresonator, and features ultra-low material-limited loss of \SI{0.15}{dB/m}~\cite{Liu2021}. We calculate the electron-photon vacuum coupling strength (numerical details in Appendix \ref{append:COMSOL}) and plot it in Fig.~\ref{fig:fig2}(b-c) as a function of electron velocity for optical wavelengths ranging from \SI{780}{\nano\meter} to \SI{2.5}{\micro\meter} (where all relevant material properties are well known), which covers most of the frequency bands that are of general interest. 

Under the optimal phase-matching condition, the interaction strength of the waveguide transverse modes scales quadratically with respect to the interaction length since they co-propagate with the electron, in contrast to the linear relation of bulk modes. In reality, waveguide-mode phase velocity differs at different optical frequencies. Through the phase-matching mechanism, linear chromatic dispersion limits the coupling bandwidths to scale inversely proportional to the interaction length. With prolonged interaction length, coupling strengths to different waveguide transverse mode families are isolated in optical frequencies, and exhibit peak features shown in Fig.\ref{fig:fig2}(b-c). Dispersion-free systems are generally feasible in higher dimensions and have been realized in specially structured photonic lattices~\cite{Vicencio15,Mukherjee15,Yang21}, where the optical modes of interest are generally unguided. In integrated photonics, advances in dispersion engineering have enabled waveguide designs that tailor the modal dispersion~\cite{Moille22,Erwan2022}, promising dispersion-free quadratic coupling enhancement over a broad frequency range. In our study, we focus on translation-invariant straight waveguides which exhibit chromatic dispersion determined by the waveguide materials and geometry. 

The waveguide mode families have finite coupling bandwidths and are well isolated from each other. We therefore define discrete spatial-temporal optical modes $\hat a_m\propto \int d\omega g_{m,\omega}\hat a_\omega$ associated with different waveguide mode families from the optical continuum based on the vacuum coupling strengths $g_{m,\omega}$ of the interaction (details see Appendix \ref{append:MD}). The coupling strength of a given mode family $\hat a_m$, 
\begin{gather}
    |g_m|^2=\int d\omega |g_{\omega,m}|^2,
\end{gather}
scales linearly with interaction length and inversely with chromatic dispersion. We quantitatively evaluate the coupling strengths $|g_m|^2$ to different spatial-temporal modes $\hat a_m$ based on the fitted interaction strength $|g_{m,\omega}|^2$ from the simulation results. As an example, for the quasi-TM\textsubscript{00} mode of the \SI{800}{nm} wide waveguide shown in Fig.\ref{fig:fig2}(c), for an electron-waveguide gap of \SI{100}{nm}, a strong coupling strength of $|g_{\mathrm{TM\textsubscript{00}}}|^2\sim1$ can be achieved with \SI{100}{\micro m} of interaction length at an electron velocity of $v_e/c=0.65$ (a kinetic energy of \SI{160}{keV}). The \SI{100}{nm} gap distance and \SI{100}{\micro m} e-beam propagation length are experimentally feasible and demonstrated in \cite{feist22} with a gradient $d|g_{\mathrm{TM\textsubscript{00}}}|^2/dz\sim\SI{5}{mm^{-1}}$.

Using the procedure described in the previous paragraph, we quantitatively investigate the influence of competing waveguide modes for a given waveguide configuration, and how one can approach unity coupling ideality by a proper choice of waveguide geometry and material, and electron beam positioning and velocity. Since the lowest order TM\textsubscript{00} mode is generally the most strongly coupled and is the most spectrally isolated mode, we target unity coupling ideality, defined by the coupling fraction 
\begin{gather}
    I \equiv |g_{\mathrm{TM\textsubscript{00}}}|^2/\int d\omega|g_\omega|^2,
\end{gather} 
to the TM\textsubscript{00} mode. 

From the numerical result shown in Fig.\ref{fig:fig2}(b-c), we find that reduced waveguide cross-section (to single-mode dimension)  enhances the mode index contrast, and results in more  spectrally isolated fundamental modes. With a better frequency isolation, the evanescent field of the coupled higher-order modes decay much faster than that of the fundamental mode in the near field, as a result of their higher optical frequencies. Therefore, one can enhance the ideality by increasing the gap distance to the waveguide surface, with $1-I$ decreasing exponentially with gap distance (details in Appendix~\ref{append:COMSOL}). 

In addition to coupling to higher-order waveguide mode families, one can also identify a rising background in the high velocity region. It can be attributed to strong coupling to the substrate modes in the Cherenkov regime ($v\gtrsim 0.7c$), where the charged particle velocity exceeds the phase velocity of light in dielectric media (here: silica). In Appendix \ref{append:substrate}, we quantify the contribution of the substrate bulk modes. This contribution can be suppressed by either choosing an electron velocity well below the Cherenkov regime of the substrate, or by using a low index material as the substrate (e.g. by suspending the structure in vacuum).

Here, we quantitatively analyze the coupling idealities in different application scenarios, and show the results in Fig.\ref{fig:fig2}(d). First, we consider state heralding applications e.g. heralded single photon sources by photon-energy loss selection with EELS. We assume an initial electron state with a fitted \SI{0.6}{eV} Voigt zero-loss-peak (ZLP) profile, and show that by conditioning on the first energy-loss sideband, one can easily achieve more than \SI{99}{\%} conditional coupling ideality $I^*$ to the TM\textsubscript{00} mode outside the Cherenkov regime ($v_e\lesssim0.7c$) with a single-mode waveguide and the electron beam positioned $\gtrsim \SI{100}{nm}$ above the surface (details in Appendix~\ref{append:COMSOL}). For a general application that is sensitive to the full optical spectrum, we show that more than \SI{95}{\%} non-conditional coupling ideality $I$ can be achieved with the electron beam placed $\gtrsim \SI{300}{nm}$ above the surface, limited by the parasitic coupling to the higher-order waveguide modes. This is not a fundamental limitation, as one can always increase the gap distance from the waveguide surface to achieve higher ideality, at the expense of reduced coupling strength. This trade-off is also illustrated in Fig.\ref{fig:fig2}(d), where the total coupling strength $|g_{\mathrm{TM\textsubscript{00}}}|^2$ is plotted against the coupling ideality. However, this effect can generally be compensated with longer interaction length $L$. As a result, given a fixed waveguide geometry and a target total coupling strength, the deviation from unity is given by $1-I \propto L^{-1}$.

In the special case where the waveguide loops and forms a resonator, the result of the open-ended waveguide studied here can equally apply (see Appendix \ref{append:res}). In most scenarios, where the electron longitudinal spatial wavefunction is shorter than the cavity round trip length, or in the frequency domain picture where the electron zero-loss-peak (ZLP) width is broader than the cavity free-spectral-range, there is no difference in terms of coupling ideality betweeen a straight waveguide and a resonator. The physical picture is that when the emitted optical pulse does not interact with the electron a second time, the emission is only determined by the local structure around the electron. In the case of a resonator, the pulse will circulate multiple times and exit the cavity as a pulse train, and exhibit in the frequency domain as a comb-structure, as was shown in~\cite{feist22,Muller2021,Auad2022}.

Note that the experimentally measured ZLP width consists of a coherent energy spread of a single-electron wavefunction, e.g. inherited from the driving laser pulses in the cold field electron emission process, and also an incoherent broadening due to e.g. the statistical imprecision of the electron acceleration voltage and the measurement instrument. In this manuscript, we mostly use ZLP width to refer to the coherent energy width, unless otherwise specified.

Generally, residual coupling to the higher-order modes can be further mitigated with heralding schemes. As an example, one can place a bandpass spectral filter~\cite{Li2021} around the frequency band of the target mode. Upon conditioning on photon-absence events at the dark port of the filter, one can further approach unity ideality, and be eventually limited by the background bulk contribution. As long as the velocity is far from the Cherenkov regime of the substrate material, we estimate this contribution to be far less than 1\%. With near-unity coupling ideality, the fidelity and purity of the interaction will be limited to the correlation between electron energy and the optical frequency components of a single spatial-temporal optical mode within the ladder state space. In the following sections, we discuss this fundamental limitation in the cases of state heralding schemes.

\begin{figure*}[t]
 	\centering
    \includegraphics[width=1\textwidth]{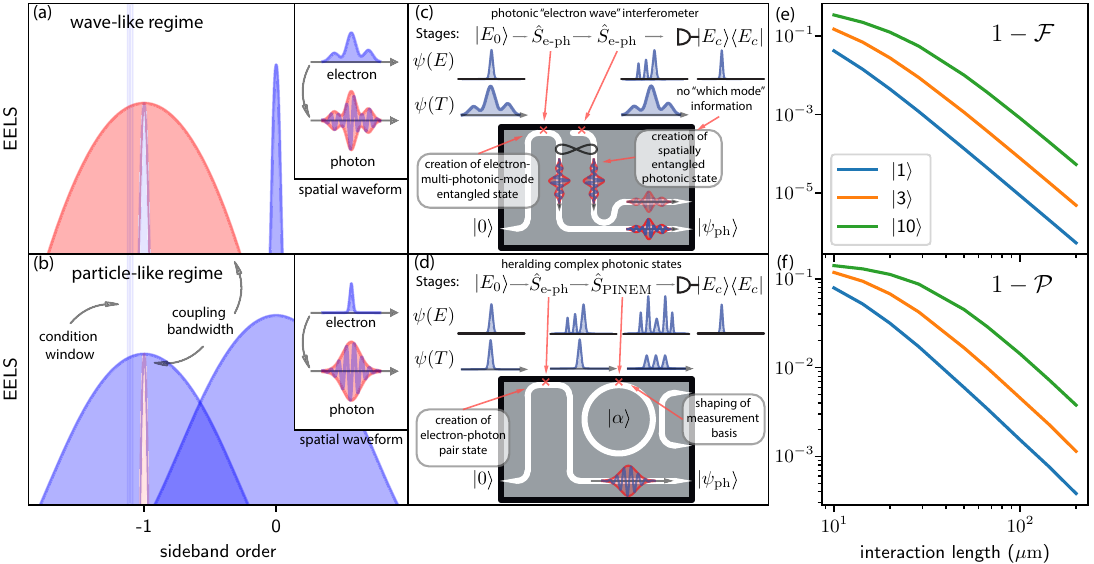}
    \caption{\textbf{(a-b)} Shaping optical waveforms by measuring electron energy. In the limit of \textbf{(a)} strong correlation with narrow zero-loss-peak (ZLP, blue) and wide phase-matching bandwidth (red), heralding results in printing electron wavefunction onto the optical waveform. In the limit of \textbf{(b)} weak correlation with wide ZLP and narrow phase-matching bandwidth, the heralded optical waveform is determined by waveguide routing and material dispersion. \textbf{(c-d)} Optical state heralding schematics where electron wavefunctions in time ($\psi(T)$) and energy ($\psi(E)$) domains are drawn before and after interaction stages (marked by red crosses).  \textbf{(c)} Scheme for electron mediated self-mode-matched optical interferometer with non-classical states, enables measurement of interferometer imbalance to the order of optical wavelength, and electron wavefunction tomography. \textbf{(d)} Scheme for heralding a general optical state by measuring electron energy, consisting of one stage for pair state preparation with $\hat S_{\mathrm{e\text{-}ph}}$ and one stage for measurement basis selection with $\hat S_{\mathrm{PINEM}}$. \textbf{(e-f)} Investigation of subspace correlation induced degradation of fidelity and purity of different Fock state components as a function of interaction length. }
    \label{fig:photon}
\end{figure*}

\section{Shaping optical states from measurement on electron energy}

In this section, we consider the case of heralding a general optical state by measuring the electron energy. To simplify the discussion here and capture the main features of the physics considered, we assume coupling to a single spatial-temporal optical mode with $I=1$, and a coherent electron wavefunction $\psi(E)$ prepared before the interaction. The effect of electron sideband overlaps (with expression shown in Appendix~\ref{appendix:optical}) is not considered since they can be efficiently eliminated experimentally and are thus not a fundamental limitation. 

We first investigate the consequences of electron-photon correlation in the state subspace in some general state preparation schemes. 
% To investigate the spatial-temporal profile of the heralded optical state, 
We consider a projection $\hat M = |E_c\rangle\langle E_c|$ on the electron's first photon sideband (general case in Appendix~\ref{appendix:optical}). This results in a pure single-photon optical state with frequency components $\phi_\omega\propto g_\omega^*\psi(E_c+\hbar\omega)$, a product between the electron wavefunction and the vacuum coupling strength. This reflects the fact that the electron energy loss is intrinsically correlated with the frequency of the photon created. The strength of the correlation depends on the initial energy uncertainty of the electron, which determines how well the photon frequency components can be distinguished by measurements of the electron state. In a stark contrast, we will see in the next section that this is not the case for the electron state heralded by photon counting, since in the no-recoil limit the frequency of the photon created does not depend on the energy of the electron. In this section, we consider two regimes of interest: The wave-like regime (section A) that exploits the correlation to its advantage, and the particle-like regime (section B) that aims for high-purity state heralding.

\subsection{Wave-like regime}
In the wave-like regime, the electron ZLP width is much narrower than the phase-matching bandwidth, as shown in Fig.~\ref{fig:photon}(a), where the electron behaves more wave-like to different optical frequency components. This regime exploits the strong correlation in the subspace between electron energy and optical frequency. This is compatible with the experimentally achieved~\cite{Krivanek19} \SI{4}{meV} ZLP width using a monochromator~\cite{Auad2022} combined with the recently demonstrated $\sim$\SI{100}{meV} phase-matching bandwidth~\cite{feist22}. In this regime, we show the expression for the heralded single-photon Fock state as
\begin{gather}
	|\psi_{ph}\rangle \propto \int   d\omega \psi(E_c+\hbar\omega)\hat a_{\omega}^\dag|0\rangle
\end{gather}
In this scenario, ignoring the waveguide dispersion during propagation, as well as electron energy dispersion, we effectively imprint the electron spatial wavefunction $\widetilde{\psi}(T=t-z/v_e)$ onto the optical waveform $\phi(T)$ of the generated single-photon Fock state of spatial-temporal mode $\hat a\propto \int d\omega \psi^*(E_c+\hbar \omega)\hat a_\omega$, with
\begin{equation}
	\phi(T) = \widetilde{\psi}(T) e^{i\omega_c T}
\end{equation}
with a center frequency $\omega_c=E_c/\hbar$ matching the conditioned electron energy $E_c$. Therefore, by shaping the electron wavefunction (e.g. pre-compression into THz pulse trains) and conditioning on a specific sideband energy, one can transfer the arbitrarily shaped electron spatial wavefunction to the optical waveform at a desired optical frequency. As for higher-order conditional Fock states $|N\rangle$, they cannot be addressed into the $N$-photon excitation of a single spatial-temporal mode (see Appendix~\ref{appendix:optical}) because the optical frequency components are highly correlated, but in any photon counting scheme, the optical profile is still shaped as $|\phi(T)|^2$ and contains $N$ photons. %since high phase-matching bandwidth usually comes with low coupling strength, we restrict ourselves to the discussion of single photon state in this regime. 

We illustrate in Fig.~\ref{fig:photon}(c) an application example in this regime. When an electron passes through two waveguides, and is then measured at the first photon sideband on the detector (single photon excitation), the measurement does not resolve in which waveguide the photons are created. In this scenario, the measurement creates a spatial entanglement of photon excitation in the two spatially separated waveguides, 
\begin{gather}
    \hat S \propto \int d\omega\psi(E_c+\hbar\omega)\left(\hat a_{1,\omega}^\dag+e^{i\omega\Delta t}\hat a_{2,\omega}^\dag\right)
\end{gather}
with naturally mode-matched waveform $\phi(T)$ and a controlled phase depending on the effective delay $\Delta t$ from the electron trajectory, essential for generating path entangled NOON states~\cite{Mitchell2004}.
If we interfere the two entangled modes with a balanced beam splitter, there is coherent quantum inference between the two waveguide excitations (see Appendix~\ref{appendix:interferometry}). In this way, we effectively constructed an optical interferometer with a non-classical optical state mediated by free electrons, with output differential photon flux
\begin{gather}
    f(t) \propto \mathrm{Re}\left[\widetilde{\psi}(t)\widetilde{\psi}^*(t+\Delta t)e^{i\omega_c\Delta t}\right].
\end{gather}
Notice that due to the nature of broadband optical coupling, when conditioning on different electron energy $E_c$, we are effectively scanning the probing optical frequency of the interferometer $\omega_c$, enabling accurate extraction of the time imbalance $\Delta t$ to the order of only a few optical cycles. When sweeping the optical path length difference to induce mode-mismatch, one can also retrieve electron spectra density based on interference visibility, similar to what was realized in matter interferometers~\cite{Hasselbach_2010}. The electron wavefunction can also be reconstructed through spectral shearing interferometry~\cite{Davis18}, answering an important question that is both fundamental and practical: how much of the measured electron energy uncertainty is quantum coherent~\cite{Aviv21}.

\subsection{Particle-like regime}
In the particle-like regime, typically associated with a long interaction length, the phase-matching bandwidth is very narrow compared to the electron energy uncertainty and the coupling strength becomes large, as shown in Fig.~\ref{fig:photon}(b). Without on-chip electron guiding structures ~\cite{Niedermayer2020,Shiloh2021}, we expect the longest interaction length to be limited to \SI{1}{mm} with $|g_\mathrm{TM_{00}}|^2\sim 5$ given beam divergence angle $\sim\SI{0.2}{mrad}$~\cite{KRIVANEK1999} with a \SI{100}{nm} gap. The electron behaves more particle-like in this regime, and can hardly distinguish different optical frequency components, therefore, the spatial-temporal optical modes defined in Section~\ref{sec:ideality} can be correctly applied. In this limit, the subspace correlation can be greatly suppressed. When conditioning on the $N$th energy sideband, we can simplify the state to photon excitations of an electron-measurement-independent spatial-temporal mode $\hat a \propto \int d\omega g_\omega \hat a_\omega$ as
\begin{gather}
    |\psi_{ph}\rangle \propto \left(\int d\omega g_{\omega}^* \hat a_{\omega}^\dag\right)^N |0\rangle,
\end{gather}
\begin{gather}
    \phi(T)\propto  \int dz \widetilde{U}_{z}^*(z,T),
\end{gather}
where the optical waveform $\phi(T)$ is connected to the Fourier transform of the optical mode profile $\widetilde{U}_{z}(z,T)=\mathrm{FT}_\omega\left[U_{z}(z,\omega)\right]$ along the electron propagation trajectory, determined by waveguide routing, and is generally much longer than the spatial extent of the electron wavefunction. For the case that includes propagation dispersion see Appendix \ref{append:state}. Since the electron travels in a straight path, by using a tailor-made waveguide structure with proper dispersion and routing, most types of optical waveforms can be generated. The center frequency of the optical excitation is not determined by the conditioned electron energy, but can be easily tuned by selecting the appropriate electron velocity, evident in the results shown in Fig.\ref{fig:fig2}(b-c). 

In the following, we restrict ourselves to the regime of long interaction length, since it is most versatile for heralding more complex optical states with higher photon numbers, and the ladder subspace correlation is weaker due to narrow phase-matching bandwidth. We show an example on how to generate highly complex optical states, with the scheme shown in Fig.~\ref{fig:photon}(d). The scheme consists of two stages, the first stage entangles the free electron with the waveguide mode, and the second stage selects the effective measurement basis for the electron energy. Specifically, the first stage of the interaction is the same pair-state generation~\cite{feist22} discussed in previous sections. While direct conditioning on the electron energy measurement generates optical Fock states, in order to generate more general optical states, one can select a more general measurement basis by having a second stage to apply a unitary transformation $\hat U$ on the electron state before the measurement. Starting from the physical measurement basis $\langle M|$, with the correct unitary transformation $\hat U$, the desired measurement basis $ \langle M| \hat U$ can be generated. If an arbitrary measurement basis can be constructed, an arbitrary quantum state can be heralded. Such a scheme exploits the time-reversal symmetry in quantum mechanics, and has been used to demonstrate optical super-resolving phase measurement using only classical lasers~\cite{Resch07}. 

In the illustrated case, shown in Fig.~\ref{fig:photon}(d), we apply a standard photon-induced near-field electron-microscopy (PINEM) operation~\cite{Henke2021}
\begin{gather}
    \hat{S}_{\mathrm{PINEM}}(g,\omega) = \exp\left(g\hat b_\omega^\dag - h.c.\right)
\end{gather}
at the same optical frequency (served as the phase reference for any follow-up optical state characterization) before detection, which effectively transforms the measurement basis from $ \langle E_c|$ to $ \langle E_c| \hat{S}_{\mathrm{PINEM}}=\sum_N c_N\langle E_c+N\hbar\omega|$ with Bessel coefficients $c_N$. Upon heralding, the generated optical state is
\begin{gather}
    |\psi_{ph}\rangle = \exp(-|g|^2/2)\sum_N\frac{c_{-(E_c/\hbar\omega+N)}g^n}{\sqrt{N!}}|N\rangle
\end{gather}
with coefficients modified by the selected electron measurement basis. Following this scheme, if at the second stage we select a more general measurement basis by modulating the electron with an optical waveform consisting of multiple harmonics~\cite{Reinhardt2020} of the base optical frequency $\hat S = \prod_n\hat S_{\mathrm{PINEM}}(g_n,n\omega)$, one can in principle generate any general optical state e.g. Cat and GKP state~\cite{Dahan22}. Note that in the no-recoil limit, any operation on the electron wavefunction commutes with the entangling operation $\hat S_{\mathrm{e\text{-}ph}}$. Therefore, it does not matter if the operation is applied post-entanglement or pre-entanglement.

Until here, we restricted ourselves to state generation in the ideal scenario where electron and photon are completely dis-entangled in the subspace of the synthetic electron-photon energy ladder. However, as is discussed in the theory section, there are still correlations between electron energy and optical frequency within the subspace. When tracing out the subspace continuum states, this leads to a degradation of state purity $\mathcal{P}=\mathrm{Tr}\left[\hat\rho^2\right]$ and fidelity $\mathcal{F}=\left|\langle\psi_{\text{prepared}}|\psi_{\text{target}}\rangle\right|^2$ of the synthesized quantum state $\hat\rho$. We analyze these effects in our state heralding scheme (expressions for $\mathcal{P}$ and $\mathcal{F}$ derived in Appendix~\ref{appendix:optical}, calculated using Monte Carlo sampling due to high dimensionality). We first stress that when conditioning on an electron energy with perfect energy resolution, the purity of the state is always unity, and we define the state fidelity in that limit. However, the relative heralding bandwidth $\gamma = \Delta E/\Delta E_{\mathrm{ZLP}}$ determines the heralding rate, and is lower-bounded by the experimental energy resolution. At finite bandwidth, it always results in finite purity of the state. We illustrate this effect at different relative heralding bandwidths in Appendix~\ref{appendix:optical}. 

We assume a ZLP width $\Delta E_{\mathrm{ZLP}}=\SI{0.6}{eV}$ with fitted Voigt lineshape from experimental data~\cite{feist22}. Given a relative heralding bandwidth $\gamma = 1$, we show in Fig.~\ref{fig:photon}(e-f) both the purity $\mathcal{P}$ of the state and the fidelity $\mathcal{F}$ compared to the target state. As the purity is only a function of the occupancy at different Fock state components, we only plot the scheme/state independent purity at these components. Due to more scrambled correlations between electron energy and photon frequency at higher ladder state subspace $|\psi_e,N\rangle$, their purity is lower, with impurity $1-\mathcal{P}\propto \sqrt{N}$. We also see that fidelity and purity increase with longer interaction distance $L$, with $1-\mathcal{F}\propto L^{-4}$ and $1-\mathcal{P}\propto L^{-2}$. This scaling is expected from the narrower phase-matching bandwidth at longer interaction length, and aligns well with the prolonged interaction targeted by the photonic integrated circuits. To help the readers grasp the inverse quadratic scaling to interaction length, we point out that for relatively short interaction length at \SI{10}{\micro\meter}~\cite{Henke2021},the state purity is $<90\%$ for Fock state components $|3\rangle$ and above, but for an interaction length at \SI{100}{\micro\meter}~\cite{feist22}, the state purity $>98\%$ even for $|10\rangle$, with fidelity exceeding $99.9\%$.

Note that any contribution from the experimental uncertainty of electron energy will lead to degradation of the electron state purity, and also increase the relative heralding bandwidth. Therefore, the experimentally measured ZLP width $\sim$\SI{0.6}{eV}~\cite{FEIST201763} can only serve as the upper bound of the quantum coherent energy uncertainty. Experimentally, the coherent energy uncertainty can be at least lower bounded at $\sim$\SI{0.1}{eV} by the measured single-electron pulse duration~\cite{BAUM201355} which is in fact still far from the Fourier limit. In the limiting case when the electron energy density matrix is completely incoherent, $\mathcal{P}\rightarrow 0$. 
%This stresses the importance of having pure particle-like electrons to generate quantum coherent optical states. 
Furthermore, as the experimentally measured purity of the heralded optical state through Wigner tomography~\cite{Leonhardt95} scales as $1-\mathcal{P}\propto \Delta E_{\mathrm{coherent}}^{-2}$, the purity characterization can also serve as a probe of the coherence property of the free electron.
 Even though the coherent electron energy width is hard to determine experimentally, it is fundamentally determined by the electron field-emission mechanism that generates the electron pulse. We can conclude that in order to be quantum coherent, the frequency spread of the heralded optical state must be much smaller than that of the laser pulses used in the electron field-emission. 

\section{Shaping electron states from optical detection}

\begin{figure}[t]
 	\centering
    \includegraphics[width=0.5\textwidth]{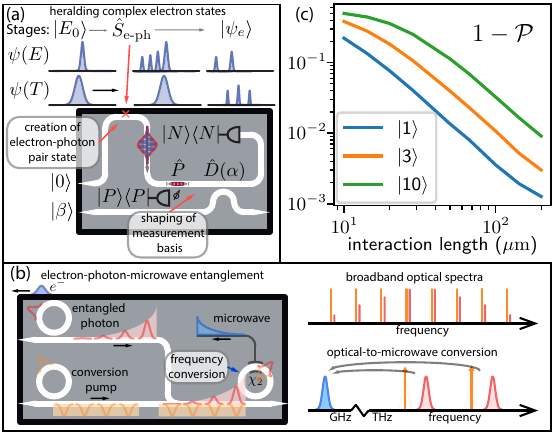}
    \caption{\textbf{(a)} Scheme for heralding an arbitrary electron state by optical detection, consisting of one stage for pair state preparation with $\hat S_{\mathrm{e\text{-}ph}}$ and one stage for measurement basis selection with on-chip optical operations. \textbf{(b)} Scheme to convert the original \SI{}{THz}-broad optical excitation to a \SI{}{MHz}-broad microwave excitation, with frequency width limited by the linewidth of the optical cavity, using a $\chi_2$ optical-to-microwave transducer. The narrow-linewidth down-converted microwave excitation is useful for interacting with \SI{}{GHz}-frequency quantum systems at low temperatures. \textbf{(c)} Investigation of subspace correlation induced degradation of purity of different electron ladder state components $|N\rangle$ as a function of interaction length. }
    \label{fig:electron}
\end{figure}

Here we consider the reciprocal operation of the previous section which is to generate complex electron energy superposition state by conditioning on photon counting. This procedure enables generation of a much broader set of electron states not accessible by conventional PINEM-type phase modulation, e.g. direct amplitude modulation of electron wavefunction. Note that with the no-recoil approximation, here the heralded spatial-temporal electron wavefunction is not shaped by the optical detection and maintains the original waveform, in sharp contrast to heralding optical state by measuring electron energy. Therefore, the fidelity $\mathcal{F}$ of the heralded electron wavefunction does not depend on interaction length, but the state purity still does. 

In Fig.~\ref{fig:electron}(a) we illustrate a similar scheme to that shown in the previous section to generate complex electron states with multiple stages of operations but on the optical side. The principle is the same, a pair-state is generated, then we select an effective measurement basis on the optical side to project the electron state into the desired form. As an example, before the detection, one can use a displacement operation $\hat D(\alpha)$, realized by a high-ratio on-chip beam splitter~\cite{Paris96}, to modify the effective photon number counting~\cite{Eaton22} measurement basis from $\langle N|$ to $\langle N| \hat D(\alpha) = \sum_{N'} c_{N'}\langle N'|$. Based on a photon counting record, a conditional electron state is prepared at
\begin{gather}
    |\psi_e\rangle =  \int dE \psi(E)\sum_N\frac{c_Ng^{N}}{\sqrt{N!}}|E-N\hbar\omega\rangle   
\end{gather}
In the special case of the coherent state measurement basis $\langle \alpha |$, which can also be constructed by simultaneously detecting both orthogonal optical quadratures in a homodyne setting, the heralding operation is equivalent to applying a direct density modulation $\exp({2|\alpha g|\cos(\frac{\omega}{v}z+\theta_\alpha)})$ on the electron wavefunction. In the limit of large modulation depth $|\alpha g|\gg 1$, the width of the electron wavefunction is compressed down to $\Delta z \sim \frac{1.7v}{\omega\sqrt{|\alpha g|}}$. The magnitude of pulse compression is similar to what is possible with PINEM-type interaction, but without the additional dispersive propagation with modulation-depth-dependent distance~\cite{Feist2015}. 

The projection into a sharply density modulated electron wavefunction by measuring in the basis of optical coherent states can be understood intuitively. Since classical coherent optical excitation can only be generated by point-like electrons, the measurement of coherent states serves as a position measurement of the electrons, projecting them into the possible periodic positions that would give the correct classical phase of the measured optical coherent state. But since the coherent states are not completely orthogonal to each other, the uncertainty of the projected electron position is determined by the magnitude of the measured field amplitude $|\alpha|$.

One can also prepare even parity electron energy state to halve the spatial modulation period, useful for generating coherent second harmonic optical emission~\cite{Kfir2021}, by applying conditional optical parity operation $\hat P(P)$ using two-level systems~\cite{Daiss19,Sipahigil2016,Najer2019} or photon-number-resolving counting~\cite{Thekkadath2020} which modifies the measurement basis to cat states $\langle \mathrm{cat}_\alpha | \propto \langle \alpha| + \langle-\alpha|$. Higher harmonic spatial modulation can be generated by detecting in higher-order cat state basis. 

On the optical side, most unitary operations or state characterizations require mode matching to a reference spatial-temporal optical mode, which is difficult to achieve for the emitted \si{\tera\hertz} broad optical pulses. Here we discuss two options that are experimentally feasible. The first option is to re-shape the emitted optical spatial-temporal profile through frequency filtering e.g. using an on-chip photonic crystal filter cavity~\cite{Sipahigil2016}. When the frequency width is narrow enough to be resolved by the detector, one can choose a continuous wave local oscillator and gate on the detector time sequence~\cite{Erwan2014} synchronized with the electron pulses. To prevent any loss-of-information that may lead to the degradation of state purity, one needs to collect all the optical excitations rejected from the filter and condition on a zero-count event from the dark port. At a single-photon level of optical emission, the relative heralding rate is determined by the filtered optical bandwidth vs. the original optical phase-matching bandwidth. Therefore, such a frequency filtering scheme does limit the heralding rate significantly due to the large phase-matching bandwidth (e.g. \SI{1}{THz} width at \SI{1}{cm} interaction length). The second option is to directly mode-match with a specifically shaped reference optical pulse. Such pulse shaping with individual control at all the frequency components is generally hard in straight waveguides, and therefore requires the use of optical resonators in place of waveguides as was recently demonstrated ~\cite{feist22}. The reference optical mode can then be generated in an identical resonator~\cite{Tikan2021} e.g. as a dissipative Kerr soliton~\cite{Kippenberg18} or an electro-optic frequency comb~\cite{Zhang2019} with control over each individual frequency component~\cite{Erwan2022,Weiner2000}. The time gating resolution required on the optical detection would then be relaxed to the optical cavity life time, which can be achieved at the level of \SI{20}{ns}~\cite{Liu2021} for materials and structures studied in the current manuscript. In integrated photonics, cavity life time approaching \SI{1}{\micro\second} is also demonstrated~\cite{Gao2022,Puckett2021,Amirhassan22}. 

Optical resonators offer the unique advantage of the concentrated optical density of states due to their narrow optical linewidth. We show a frequency conversion example in Fig.~\ref{fig:electron}(b) (details of the scheme see Appendix~\ref{append:conversion}) to exploit this advantage of optical resonators to convert the \SI{}{THz}-broad optical excitation from the electron-photon interaction to a \SI{}{MHz}-broad microwave excitation using a $\chi_2$ optical-to-microwave converter~\cite{Sahu2022,Fan18}. Using a structured local oscillator pump field, the conversion effectively serves as a multi-mode demodulation of the entangled photons. The frequency width of the microwave photon is determined by the linewidth of the optical cavity mediating the electron-photon interaction. Compared to the original \SI{}{THz}-broad optical excitation, this frequency conversion is particularly useful to bridge interactions of \SI{}{eV}-broad free electrons with quantum systems at \SI{}{GHz} frequencies, e.g. superconducting qubits, electron spin qubits and mechanical oscillators. Generally, with the coupling to well-controlled two-level systems in the strong coupling regime, any photon measurement basis can be constructed~\cite{Law96}. As arbitrary quantum state synthesis of microwave photons was experimentally demonstrated in superconducting qubit systems \cite{Hofheinz2009}, we can construct an arbitrary measurement basis $\langle \psi| = \langle 0|\hat U$ by applying the unitary operation $\hat U$ on the converted microwave field and then conditioning on the microwave ground state $\langle 0 |$ with photon-number resolving measurement~\cite{Schuster2007} using a superconducting qubit, promising arbitrary electron state generation. Optical-to-microwave converters and superconducting qubits mostly require \SI{}{mK} temperatures due to their \SI{}{GHz}-frequencies, and usually operate in a dilution refrigerator. Therefore, optical excitations need to be guided out of the electron microscope through optical fibers, stressing the importance of high efficiency fiber-to-chip couplings~\cite{Almeida03}.

Note that due to multiple stages, usually the optical measurement event will occur after the electron detection due to the high electron velocity. However, a delayed measurement on the optics side does not impair our scheme, as the measurement operators on the two parties commute~\cite{Kim2000}. Therefore, no real-time action is required.

Here, we show the full bandwidth state purity (expressions derived in Appendix~\ref{appendix:electron}) as a function of interaction length in Fig.~\ref{fig:electron}(c). As expected, it follows the same $1-\mathcal{P}\propto L^{-2}\sqrt{N}$ scaling, and favors longer interaction length. We point out again that for a relatively short interaction length at \SI{10}{\micro\meter}, the electron ladder $|10\rangle$ state  purity is $50\%$, but for an interaction length at \SI{200}{\micro\meter}, the state purity reaches $99\%$. Note that the state purity is completely determined by the electron-photon interaction, and does not depend on specific schemes e.g. optical-to-microwave conversion.

Here, a lower purity of the initial electron state will also result in purity degradation of the heralded electron wavefunction, similar to the case of heralded optical state discussed in the previous section. However, effects like heralded density modulation is robust as the electron position projections are always valid given optical measurement records even with mixed electron states.

\section{Discussion}
We analyzed fundamental limits of integrated photonic circuits as a platform for synthesizing high-quality quantum states with free electrons. We show that near-unity coupling ideality to the target TM\textsubscript{00} spatial-temporal waveguide mode can be achieved by suppressing parasitic couplings through the control of electron beam positioning, velocity, and waveguide design. We also investigate the underlying correlation between electron energy and photon frequency in the energy ladder subspace, and the induced fundamental limitation as a trade-off between heralding rate and state purity. We found that particle-like electrons with coherent energy uncertainty are required to generate pure heralded states, and the purity limit can be greatly relaxed with experimentally feasible interaction length with integrated waveguides. We also show that these correlations can be exploited to shape the optical waveforms, e.g. to map the electron wavefunction to optical domain and construct an effective optical interferometer mediated by free electrons. However, the spatial sensitivity of such an interferometer remains at the optical-wavelength scale, and does not inherit the superior spatial sensitivity of electron waves. It is still an open question whether phase-object-induced electron phases can be transferred to the optical domain, accumulate and be detected optically, which is relevant for quantum-enhanced phase-object imaging applications~\cite{KRUIT2016, Okamoto22}. We also anticipate that the maximum feasible coupling strength $|g_\mathrm{TM_{00}}|^2$ can be further enhanced through waveguide dispersion engineering~\cite{Moille22}, with the trade-off of lower state purity due to larger phase-matching bandwidth.

Note that in our discussion we omit detailed analysis on some experimental limitations e.g. finite detection efficiencies, primarily on the optical side. The heralded optical state is robust given the high energy of the electrons, but the heralded electron state purity is therefore most sensitive to the optical detector efficiencies and other limiting factors such as optical mode-matching. We anticipate that there are schemes or parameter regimes that are less prone to detection inefficiencies. We also restrict our discussion mostly to the interaction picture, except that the electron and optical waveforms are defined in the Schr\"{o}dinger picture. We remind the reader that in the Schr\"{o}dinger picture, though not the main focus of the paper, long distance propagation significantly modifies the electron and optical density profile $|\widetilde{\psi}(T)|^2$ and $|\phi(T)|^2$, leading to effects like electron~\cite{Talebi2018,Sergey21,Priebe2017,Morimoto2018,Norbert19} and optical~\cite{Tomlinson84} pulse compression. 
Moreover, in the no-recoil limit, all the electron operations commute with each other. This approximation, though practically valid for few-photon single-chip interactions, limits the controllable degrees of freedom of the heralded optical states to the order of the harmonics of the PINEM field used to shape the electron wavefunction~\cite{Reinhardt2020}. In the platform of \SiN microresonators, efficient generation of second~\cite{Nitiss2022} and third~\cite{Wang16,Siddharth22} harmonics are supported with estimated maximum $|g_2|\sim100$ and $|g_3|\sim10$, offering a total of $8$ degrees of freedom on the heralded optical state. The \SiN{} integrated photonics platform also provides an ultra-wide transparency window from \SI{400}{nm} to \SI{4.5}{\micro\meter}~\cite{Tran2022}, supporting at most ten harmonic components with an externally driven optical source. Beyond the no-recoil limit~\cite{Talebi2018}, electron energy transitions significantly modifies the velocity due to energy dispersion. When interaction regions are far apart, the recoil effect results in an energy-dependent phase accumulation between different stages (details see Appendix~\ref{appendix:recoil}). The recoil effect can be safely neglected in the discussion of few-photon single-chip interaction, but could be important for a wider range of experimental schemes~\cite{Yalunin2021,Kfir2021,Priebe2017}, e.g. when multiple chips are involved with significant separation distance.  
Furthermore, our analysis is purely in the framework of macroscopic QED\cite{3dquant}, where electrons interacts with the medium-assisted electromagnetic fields. In optical media like the \SiN material we discuss here, optical phonons typically exist and result in Raman scattering of optical fields~\cite{Karpov16}. But, due to their short spatial extent and low energy~\cite{Krivanek2014,Venkatraman2019}, we do not consider their contribution in long-distance phase-matched interaction with the high-energy free electrons. The same reasoning also excludes higher electron energy loss process e.g. valence and inner-shell ionization around \SI{50}{eV}~\cite{Hofer_2016}.

Our analysis and results indicate that the photonic integrated circuit platform is ideal for free-electron quantum optics with manageable limitations, and promises a pathway to high-fidelity and high-purity quantum state heralding, entanglement of free electrons with other quantum systems, and quantum-enhanced sensing and imaging.

\section*{Acknowledgments}
We thank Armin Feist, Germaine Arend, Thomas Juffmann, Xihang Shi, Yujia Yang and Terence Bl\'{e}sin for useful discussions. This work was supported by the Swiss National Science Foundation under grant agreement 185870 (Ambizione). O.K. gratefully acknowledges the Young Faculty Award from the National Quantum Science and Technology program of the Israeli Planning and Budgeting Committee.

%%%%%%%%%%%%%%%%%%%%%%%%%%%%%%%%%%%%%%%%%%%%%%%%%%%%%%%%%%%%%%%%%%%%%%%%%%%%
%%%%%%%%%%%%%%%%%%%%%%%%%%%%%%%% Appendices %%%%%%%%%%%%%%%%%%%%%%%%%%%%%%%% 
%%%%%%%%%%%%%%%%%%%%%%%%%%%%%%%%%%%%%%%%%%%%%%%%%%%%%%%%%%%%%%%%%%%%%%%%%%%%

\appendix

\counterwithin{figure}{section}

\section{QED details }\label{appendix: theory}

We consider an electron beam with a narrow momentum spread around wavevector $\bd{k}_0$ and assume that the photon energies involved in the interaction are much smaller than the electron relativistic energy $E_0 = c\sqrt{c^2m^2+\hbar^2 k_0^2}$ (i.e. the no-recoil regime). In the velocity gauge, the Hamiltonian is described as~\cite{Kfir2021,DiGiulio2021}
\begin{gather*}
    \hat H = \hat H_{\mathrm{el}} + \hat H_{\mathrm{ph}} + \hat H_{\mathrm{int}}\\
    \hat H_{\mathrm{el}} = \sum_{\bd{k}} [E_0 + \hbar \bd{v}\cdot(\bd{k}-\bd{k}_0)]\hat c_{\bd{k}}^\dag \hat c_{\bd{k}}\\
    \hat H_{\mathrm{ph}} = \int d\omega \int d^3\bd{r}\hbar\omega \hat{\bd{f}}^\dag(\bd{r},\omega)\cdot \hat{\bd{f}}(\bd{r},\omega)\\
    \hat H_{\mathrm{int}} = -\int d^3\bd{r}\hat{\bd{J}}(\bd{r})\cdot \hat{\bd{A}}(\bd{r})
\end{gather*}
where we defined the electron current operator $\hat{\bd{J}}(\bd{r})=(-e\bd{v}/V)\sum_{\bd{k},\bd{q}}e^{i\bd{q}\cdot\bd{r}}\hat c_{\bd{k}}^\dag\hat c_{\bd{k}+\bd{q}}$ using the Fermionic ladder operators $\hat c_{\bd{k}}$ and the relativistic electron group velocity $\bd{v} = \hbar c^2 \bd{k}_0/E_0$, and a linear electron energy dispersion is assumed. The vector potential $\hat{\bd{A}}(\bd{r},t)=\int \frac{d\omega}{2\pi}\hat{\bd{A}}(\bd{r},\omega)e^{i\omega t} + h.c.$ is associated with the noise current operator $\hat{\bd{j}}^{\mathrm{noise}}(\bd{r},\omega)$ through the quantized three-dimensional Maxwell equation\cite{3dquant} and has a formal solution
\begin{gather*}
    \hat{\bd{A}}(\bd{r},\omega) = -4\pi \int d^3\bd{r}'G(\bd{r},\bd{r}',\omega)\cdot\hat{\bd{j}}^{\mathrm{noise}}(\bd{r}',\omega)
\end{gather*}
where $G(\bd{r},\bd{r}',\omega)$ is the dyadic Green function (Green tensor) of the classical problem satisfying the equation
\begin{gather*}
    \nabla\times\nabla\times G(\bd{r},\bd{r}',\omega) - \frac{\omega^2}{c^2}\epsilon(\bd{r},\omega)G(\bd{r},\bd{r}',\omega)=-\mu_0\delta(\bd{r}-\bd{r}')
\end{gather*}
which describes the field response at $\bd{r}$ to a point current excitation at $\bd{r}'$. Since we are dealing with non-magnetic materials, we assume a relative permeability $\mu(\bd{r})=1$. The noise operator is bosonic and was chosen to be
\begin{gather*}
    \hat{\bd{j}}^{\mathrm{noise}}(\bd{r},\omega)=\omega\sqrt{\hbar\epsilon_0\mathrm{Im}\{\epsilon(\bd{r},\omega)\}}\hat{\bd{f}}(\bd{r},\omega)
\end{gather*}
in order to satisfy the fluctuation-dissipation theorem due to dissipative material, with bosonic ladder operators $\hat{\bd{f}}(\bd{r},\omega)$ satisfying commutation relation $\left[\hat{f}_i(\bd{r},\omega), \hat{f}_i'(\bd{r}',\omega')\right]=\delta_{i,i'}\delta(\bd{r}-\bd{r}')\delta(\omega-\omega')$. Note that in the limiting case of a dispersive material (assumed in this study, characterized by its instantaneous electronic response) $\mathrm{Im}\{\epsilon(\bd{r},\omega)\}\rightarrow 0$. However, this imposes no problem for our formalism which is shown to correctly reduce to the mode decomposition method used in the quantized vacuum field\cite{Gruner1995} due to Kramers–Kronig relations. 

When projecting to the direction of the electron trajectory $\hat{\bd{z}}$ with transverse coordinate $\bd{R}_0$, the scattering matrix is shown to be
\begin{gather*}
    \hat S = e^{i\hat\chi}\hat U\\
    \hat U = \exp\left\{\left[\frac{-ie}{2\pi\hbar  V^{2/3}}\sum_{\bd{k},\bd{q}_\perp}\int d\omega \int d^3 \bd{r} e^{i\bd{q}_\perp\cdot \bd{R}}e^{-i\omega z/v_e}\right.\right.\\
    \left.\left.\hat{A}_z(\bd{r},\omega)\hat c_{\bd{k}}^\dag \hat c_{\bd{k}+\bd{q}_\perp-(\omega/v)\hat{\bd{z}}}\vphantom{\sum_{1,2}}\right]-h.c.\right\}
\end{gather*}
where $\bd{q}_\perp$ is the transverse component of the exchanged electron wave vector. We can further simplify the expression by disregarding the phase operator $\hat \chi$ and assuming a point electron distribution over the transverse direction, and obtain
\begin{gather*}
    \hat U = \exp\left[\int d\omega g_\omega\hat b_\omega^\dag \hat a_\omega-h.c.\right]
\end{gather*}
where continuum photon and electron operators are introduced
\begin{gather*}
    \hat a_\omega = -\frac{ie}{2\pi\hbar g_\omega}\int dz e^{-i\omega z/v_e}\hat A_z(\bd{R}_0,z,\omega)\\
    \hat b_\omega = \sum_{k_z}\hat c_{k_z}^\dag\hat c_{k_z+\omega/v}
\end{gather*}
with vacuum coupling strength $g_\omega$ associated with the electron energy loss (EELS) probability studied in this manuscript
\begin{gather*}
    |g_\omega|^2 = \Gamma(\bd{R}_0, \omega) = \frac{4e^2}{\hbar}\iint dz dz'\\ \mathrm{Re}\{ie^{i\omega(z-z')/v}G_{zz}(\bd{R}_0,z;\bd{R}_0,z';\omega)\}.
\end{gather*}
The operators are defined in this way so that the quantum optical commutation relations are preserved $[\hat a_\omega,\hat a_{\omega'}^\dag]=\delta(\omega-\omega')$, and can be easily proven using the identity
\begin{gather*}
    \sum_{i''}\int d^3\bd{r}''\mathrm{Im}\{\epsilon(\bd{r}'',\omega)\}G_{i,i''}(\bd{r},\bd{r}'',\omega)G_{i',i''}^*(\bd{r}',\bd{r}'',\omega)\\
    =-\frac{1}{\epsilon_0\omega^2}\mathrm{Im}\{G_{i,i'}(\bd{r},\bd{r}',\omega)\}.
\end{gather*}
Note that $\hat a_\omega$ contains contributions from all the spatial modes at $\omega$, and is not a specific pre-defined spatial mode $\hat a_{\omega,m}$ which is frequently used in cavity QED systems. 

To find the spatial mode function of $\hat a_\omega$, we use the following relations~\cite{Johannes21} for an arbitrary set of orthogonal basis $\hat a_{i,\omega}$
\begin{gather*}
    \hat a_{i,\omega} = \int d^3\bd{r}\bd{V}_i(\bd{r},\omega)\cdot\hat {\bd{f}} (\bd{r},\omega)\\
    \hat{\bd{f}} (\bd{r},\omega) = \sum_i\bd{V}_i^\dag(\bd{r},\omega)\hat a_{i,\omega}
\end{gather*}
where the weight functions obeys the following normalization condition
\begin{gather*}
    \int d^3\bd{r}\bd{V}_i(\bd{r},\omega)\cdot\bd{V}_j^\dag(\bd{r},\omega) = \delta_{i,j}.
\end{gather*}
From here, we can re-express the field operator in terms of the set of orthogonal basis as
\begin{gather*}
    \hat{\bd{A}}(\bd{r},\omega) = -4\pi \omega\int d^3\bd{r}'\sqrt{\hbar\epsilon_0\mathrm{Im}\{\epsilon(\bd{r}',\omega)\}}G(\bd{r},\bd{r}',\omega)\\
    \cdot\sum_i\bd{V}_i^\dag(\bd{r}',\omega)\hat a_{i,\omega}.
\end{gather*}
By choosing one of the spatial modes $\hat a_{i=0,\omega}$ as our mode of interest $\hat a_{\omega}$ with the weight function
\begin{gather*}
    \bd{V}_{\hat a_{\omega}}(\bd{r},\omega) = -\frac{2ie\omega}{g_\omega}\sqrt{\frac{\epsilon_0}{\hbar}}\int dz e^{-i\omega z/v_e}\\
    \int d^3\bd{r}'\sqrt{\mathrm{Im}\{\epsilon(\bd{r}',\omega)\}}\hat{\bd{z}}\cdot G(\bd{R}_0,z;\bd{r}';\omega),
\end{gather*}
we can find the mode function of this optical mode as
\begin{gather*}
    \hat {\bd{A}}_{\hat a_\omega}(\bd{r},\omega) = 2\pi\sqrt{\frac{\hbar}{2\omega\epsilon_0}}\bd{U}_{\hat a_\omega}(\bd{r},\omega)\hat a_\omega \\
    \bd{U}_{\hat a_\omega}(\bd{r})=\frac{-4e}{g_\omega^*}\sqrt{\frac{2\epsilon_0\omega}{\hbar}}\int dz e^{i\omega z/v_e}\mathrm{Im}[G(\bd{r};\bd{R}_0,z;\omega)\cdot\hat{ \bd z}],
\end{gather*}
which is a mode specifically defined for this interaction. This mode construction corresponds to a linear transformation of the original structure-supported optical spatial modes, such that only this specific optical mode is involved in the interaction, while all the other transformed modes are dark and invisible to the electron.
It is in this way advantageous to use this formalism to account for the infinite number of spatial modes of the optical structure that the electron couples to. In the limit of unity coupling ideality, this mode function converges to the one of the waveguide modes. If the electron transverse spread is significant, the EELS probability is shown \cite{Ritchie1988} to be an average over the transverse electron wavefunction
\begin{gather*}
    \Gamma( \omega)=\int d^2\bd{R}|\psi_\perp(\bd{R})|^2\Gamma(\bd{R}, \omega).
\end{gather*}
However, this type of averaging is not quantum coherent. At different transverse positions, the coupling coefficients are different. Therefore, we have to modify the scattering matrix to
\begin{gather*}
    \hat S = \exp\left[\int d\omega d^2\bd{R} g_\omega(\bd{R})\hat b_\omega^\dag|\bd{R}\rangle\langle\bd{R}| \hat a_\omega-h.c.\right].
\end{gather*}
If the part of the longitudinal optical field that overlaps with the electron transverse wavefunction has considerably inhomogeneity, then the different transverse position components of the electron will be entangled with different longitudinal electron-photon pair states, characterized by their different coupling strengths. Therefore, information loss occurs when tracing out the transverse degrees of freedom of the electron, leading to state purity degradation. Since \SI{}{nm}-scale electron beam focuses are routinely used in electron microscopes, this is not a significant limitation for near-field coupling to optical waveguides which have a typical decay length of $\sim$\SI{100}{nm}. 

\section{Equivalence to the classical result}\label{append:class}

The electron energy loss at a dielectric surface can be interpreted classically in a microscopic picture~\cite{Ritchie1957}, see Fig.\ref{fig:illus}: if an electron passes near the surface of a dielectric structure, the dipoles in the structures are polarized (equivalently classical current), induced by the electric field from the flying electron, and generates a backaction field $\bd{E}(\bd{r}_e(t), t)$ to the electron at $\bd{r}_e(t)$ that induces electron energy loss. The total energy loss can be expressed in time domain and frequency domain as
\begin{equation*}
	\Delta E = e \int dt \bd{v}\cdot\bd{E}(\bd{r}_e(t), t) = \int \hbar\omega d\omega \Gamma(\omega)
\end{equation*}
where the frequency domain energy loss function $\Gamma(\omega)$ is expressed as
\begin{equation*}
	\Gamma(\omega) = \frac{e}{\pi\hbar\omega}\int dt \mathrm{Re}\left[e^{-i\omega t}\bd{v}\cdot\bd{E}(\bd{r}_e(t), \omega)\right]
\end{equation*}
which can be easily verified if one plugs it back into the energy loss expression and the correct time integral is retrieved. Notice that here $\bd{E}(\bd{r}_e(t), \omega)$ is not the direct Fourier transform of $\bd{E}(\bd{r}_e(t), t)$. The Fourier transform applies only on the time dependence of the electric field function not explicitly depending on the electron trajectory function $\bd{r}_e(t)$. The frequency domain components depends explicitly on the current induced from a given electron trajectory, but do not take into account the sampling of the field at different position $\bd{r}_e(t)$ at different time $t$. This ensures that the total energy loss is consistent, but renders the formalism non-local. This treatment is consistent with the quantum formalism when the electron is decomposed into perfect momentum states where the wavepacket length is infinite, as one could see from the fact that even though the electron only interacts with the structure locally, the resulting energy loss spectrum will show e.g. discrete mode structure (a non-local property). This is the result of this particular Fourier expansion procedure, but when considering the electron in terms of wave packets this treatment is valid. It has been shown~\cite{Ritchie1988} that a full quantum treatment gives exactly the same EELS result.

Using the no-recoil approximation, which assumes that the radiation of electron into the surrounding substrates does not change the trajectory $\bd{r}_e(t)$ of the electron significantly, we can directly calculate the induced electric field $\bd{E}(\bd{r}_e(t), t)$ from the electron current $\bd{j}(\bd{r}, t)$ through the Green tensor of the whole dielectric structure,
\begin{equation*}
	\bd{E}(\bd{r}, \omega) = -4\pi i\omega\int d^3\bd{r}'G(\bd{r},\bd{r}',\omega)\cdot\bd{j}(\bd{r}', \omega )
\end{equation*}
where the Green tensor $G(\bd{r},\bd{r}',\omega)$ is the elementary solution of the full Maxwell equation 
\begin{equation*}
	\nabla\times\nabla\times G(\bd{r},\bd{r}',\omega) - \frac{\omega^2}{c^2}\epsilon(\bd{r},\omega)G(\bd{r},\bd{r}',\omega)=-\mu_0\delta(\bd{r} - \bd{r}')
\end{equation*}
with a point current at position $\bd{r}'$ in frequency domain. A flying electron is equivalent to a broadband evanescent source, and here we consider an electron beam at $\hat{\bd{z}}$ direction at transverse coordinate $\bd{R}_0$, for which the frequency domain electron current density is
\begin{equation*}
	\bd{j}(\bd{r}, \omega ) =-e\hat{\bd{z}}\delta(\bd{R}-\bd{R}_0)e^{i\omega(z-z_0)/v}.
\end{equation*}
From here, one can express the frequency domain loss rate in terms of Green function as
\begin{equation*}
	\Gamma(\omega) = \frac{4e^2}{\hbar}\int dz dz' \mathrm{Re}[ie^{i\frac{\omega(z-z')}{v}}G_{zz}(\bd{R}_0,z;\bd{R}_0,z';\omega)],
\end{equation*}
which coincides with the result from a full QED treatment. One should keep in mind that the Green tensor here has two contributions, one from vacuum $G_0$ when there is no structure around, and the other component from the backaction field $G_\mathrm{ind}$ that is induced from the dielectric dipoles. Only the backaction field $G_\mathrm{ind}$ contributes to electron energy loss, because electron does not emit in vacuum so the contribution from the vacuum $G_0$ vanishes in the integral.

\section{Correspondence to modal decomposition }\label{append:MD}

The correspondence between the 3D macroscopic quantization method in a dispersive material with the conventional quantum optics quantization procedure using modal decomposition has been demonstrated for the 1D case~\cite{Gruner1995}.  Here, we show the correspondence with the quantum optical formalism used in~\cite{Henke2021}. To account for all the spatial modes at a given frequency $\omega$, the quantization of vector potential was chosen as
\begin{gather*}
    \hat{\bd{A}}(\bd{r},\omega) = -4\pi \omega\int d^3\bd{r}'\sqrt{\hbar\epsilon_0\mathrm{Im}\{\epsilon(\bd{r}',\omega)\}}G(\bd{r},\bd{r}',\omega)\cdot\hat{\bd{f}}(\bd{r}',\omega)
\end{gather*}
to fulfill the canonical field commutation relations. However, in vacuum or lossless media, the modal decomposition method~\cite{Roy91} is often used instead, with
\begin{gather*}
    \hat{\bd{A}}(\bd{r},t) = \sum_m \sqrt{\frac{\hbar}{2\omega_m \epsilon_0}}\bd{U}_m(\bd{r})\hat{a}_{\omega_m,m}e^{-i\omega_m t}+h.c.
\end{gather*}
where the profile function $\bd{U}_m(\bd{r})$ of each mode defined in a frequency window $\Delta\omega_m$ satisfies the wave equation
\begin{gather*}
    \nabla\times\nabla\times \bd{U}_m(\bd{r}) - \frac{\omega^2}{c^2}\epsilon(\bd{r},\omega_m)\bd{U}_m(\bd{r})=0
\end{gather*}
with normalization condition
\begin{gather*}
    \int d^3\bd{r}\epsilon(\bd{r},\omega_m)\bd{U}_m(\bd{r})\cdot\bd{U}_n^*(\bd{r}) = \delta_{m,n}.
\end{gather*}
From here, one can easily find the correspondence between the spatial mode ladder operators $\hat{a}_{\omega,m}$ and the bosonic ladder operators $\hat{\bd{f}}(\bd{r}',\omega)$ as
\begin{gather*}
    \hat{a}_{\omega_m,m} =  -4\pi \int_{\Delta\omega_m} d\omega\iint d^3\bd{r}d^3\bd{r}'\\
    \sqrt{2\omega^2\omega_m\mathrm{Im}\{\epsilon(\bd{r}',\omega)\}}\epsilon_0\epsilon(\bd{r},\omega)\bd{U}_m^*(\bd{r})\cdot G(\bd{r},\bd{r}',\omega)\cdot\hat{\bd{f}}(\bd{r}',\omega),
\end{gather*}
with their vacuum coupling strength to the electron as~\cite{feist22} 
\begin{gather*}
    g_{\omega_m,m}(\bd{R}_0) = -i\sqrt{\frac{e^2}{2\epsilon_0\hbar\omega_m }} \int dze^{-i\omega_m z/v_e}U_{m,z}(\bd{R}_0,z).
\end{gather*}

In this formalism, we can rewrite the scattering matrix in its modal decomposition form
\begin{gather*}
    \hat S = e^{i\hat \chi}\exp\left[\int d\omega g_\omega\hat b_\omega^\dag \hat a_\omega-h.c.\right]\\
    = e^{i\hat \chi'}\exp\left[\sum_m g_{\omega_m,m}\hat b_{\omega_m}^\dag \hat a_{\omega_m,m}-h.c.\right]
\end{gather*}
where the optical mode operators $\hat a_{\omega_m,m}$ are no longer continuum mode operators and now satisfy $[\hat a_{\omega_m,m},\hat a_{\omega_n,n}^\dag]=\delta_{m,n}$. 

In the case of an optical cavity, the optical modes are well-defined bosonic modes. As long as the electron energy resolution does not resolve frequency components of the optical mode, the treatment is valid. For an open waveguide, the modes that are coupled to the electron are instead travelling modes in a continuum~\cite{continuum}. This is the most general case and can include the open cavity modes as well.
The vacuum coupling strength of a continuum frequency mode in a spatial mode family is
\begin{gather*}
    g_{\omega,m}(\bd{R}_0) = -i\sqrt{\frac{e^2}{2\epsilon_0\hbar\omega }} \int dze^{-i\omega z/v_e}\tilde{U}_{m,z}(\bd{R}_0,z,\omega),
\end{gather*}
where the profile function $\tilde{\bd{U}}_m(\bd{r},\omega)$ satisfies the wave equation as well, but with normalization condition
\begin{gather*}
    \int d^3\bd{r}\epsilon(\bd{r},\omega)\tilde{\bd{U}}_m(\bd{r},\omega)\cdot\tilde{\bd{U}}_n^*(\bd{r},\omega') = \delta_{m,n}\delta(\omega,\omega').
\end{gather*}
Index $m$ here represents different spatial mode families. However, when the electron energy resolution does not resolve the frequency structure of the coupling strength to any given mode family, as in the \emph{weak correlation regime} discussed in the manuscript, one can still define the corresponding non-continuous operators for different mode families,
\begin{gather*}
    \hat{a}_m = \int d\omega\phi_m^*(\omega)\hat{a}_{\omega}\\
    \hat{a}_{\omega} = \sum_m\phi_m(\omega)\hat{a}_m
\end{gather*}
where $\phi_m(\omega)$ is the Fourier component of the temporal field profile functions~\cite{Brecht15,Raymer_2020}
\begin{gather*}
    \phi_m(\bd{r},t)=i\int d\omega\sqrt{\frac{\hbar\omega}{2\epsilon_0}}\phi_m(\omega)\tilde{\bd{U}}_m(\bd{r},\omega)e^{-i\omega t}\\
    \hat{E}(\bd{r},t) = \sum_m \phi_m(\bd{r},t)\hat{a}_m + h.c.
\end{gather*} 
of the defined mode families. It is a complete orthogonal set of functions on $\omega$,
\begin{gather*}
    \int d\omega \phi_m(\omega)\phi_n^*(\omega)=\delta_{m,n}\\
    \sum_m\phi_m(\omega)\phi_m^*(\omega')=\delta(\omega-\omega')
\end{gather*}
found through Gramm-Schmit orthonormalization procedure, such that the commutation relation $[\hat{a}_m,\hat{a}_n^\dag]=\delta_{m,n}$ is satisfied for these \emph{field operators} in the context of quantum field theory, introduced to avoid using operator-valued distributions. One can therefore rewrite the scattering matrix in the new mode family field operator basis
\begin{gather*}
    \hat S = e^{i\hat \chi}\exp\left[\sum_mg_m\hat b_m^\dag \hat a_m-h.c.\right]
\end{gather*}
where $g_m = \int d\omega g_\omega\phi_m(\omega)$. The total coupling strength would be $|g_m|^2 = \iint d\omega d\omega' g_\omega g_{\omega'}^*\phi_m(\omega)\phi_m^*(\omega')$. Here, when the frequency bands of different mode families with non-negligible coupling strength $g_\omega$ are sufficiently separated, we choose the profile function $\phi_m(\omega)=\mathbb{I}_{\omega\in \Delta\omega_m}g_\omega^*/g_m^*$, where ${\Delta\omega_m}$ is the frequency window within which we define the field operator for the corresponding mode family, and $|g_m|^2 = \int_{\Delta\omega_m} d\omega |g_\omega|^2$. Note that when the coupling to bulk modes is significant, one has to use the coupling strength $g_{\omega,m}$ from the conventional modal decomposition method instead of the Green function method to quantitatively isolate the coupling to a mode family from background bulk mode contributions.

The cavity mode decomposition is actually the narrow-band approximation of the Gramm-Schmit orthonormalization procedure, where $\phi_m(\omega)$ is strongly peaked around the mode center frequency, since all optical modes, though narrow, still have a finite linewidth due to the coupling to outside channels (e.g. bus waveguide and cavity losses). The profile function can be found through the input-output formalism~\cite{Gardiner85} of an optical cavity $\hat a_m$, assuming unity coupling efficiency to the bus waveguide mode $\hat a_{\mathrm{out}}$ with coupling rate $\kappa$,
which results in a profile function of $\phi_m(\omega) \propto \frac{\sqrt{\kappa}}{-\frac{\kappa}{2}+i(\omega_m-\omega)}$, where the bus waveguide is part of the resonator and forms the continuum modes in frequency domain. 

\section{COMSOL simulation details}\label{append:COMSOL}
\begin{figure*}[t]
 	\centering
    \includegraphics[width=1\textwidth]{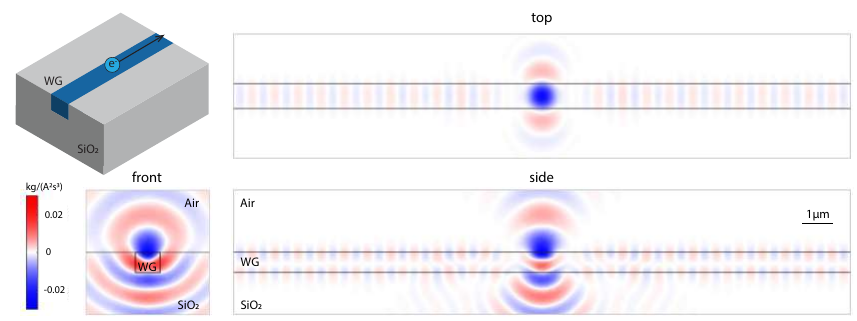}
    \caption{ 
    Spatial pattern of $\mathrm{Im}[G_{zz}(\bd{r},\bd{r}_0,\omega)]$ for the case of a \SiN waveguide embedded in a silica substrate. In addition to emission into the substrate and free space, some guided modes in the waveguide are also excited by the oscillating electric current dipole, and forms a beating spatial pattern amongst guided modes along the waveguide direction. 
    }
    \label{fig:Green}
\end{figure*}

\begin{figure}[t]
 	\centering
    \includegraphics[width=0.5\textwidth]{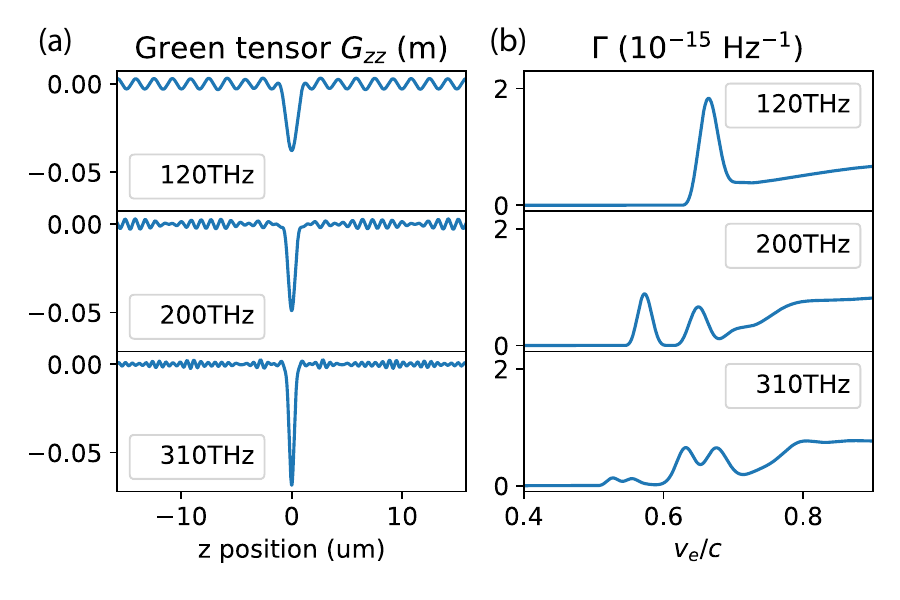}
    \caption{
    \textbf{(a)} Examples of the Green function $\mathrm{Im}[G_{zz}(\bd{r},\bd{r}_0,\omega)]$ along the trajectory of the electron at different optical frequencies, and \textbf{(b)} the corresponding vacuum coupling strength at different electron velocities, for a \SI{50}{\micro\meter} interaction length. The spatial beating of many mode families is visible in the Green functions, and also in the coupling strength. The coupling to different mode families is phase-matched at different electron velocities at a given optical frequency. When the electron velocity is in the Cherenkov regime ($v\gtrsim 0.7c$), the energy loss is eventually dominated by the substrate loss. }
    \label{fig:Green_APP}
\end{figure}

\begin{figure*}[t]
 	\centering
    \includegraphics[width=1\textwidth]{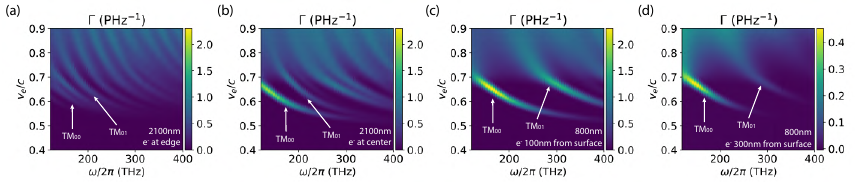}
    \caption{ Electron-photon coupling spectrum with \SI{50}{\micro\meter} interaction length for different waveguide geometries and electron positioning. The coupling spectrum is plotted as a function of both electron velocity $v_e$ and optical frequency $\omega$. The waveguides have a thickness of \SI{650}{nm}, and widths of \textbf{(a-b)} \SI{2.1}{\micro m} and \textbf{(c-d)} \SI{800}{nm}, and are embedded in a silica substrate. Coupling to different waveguide mode families appears as multiple coupling bands. Coupling ideality to the target TM\textsubscript{00} mode is improved by changing the electron beam transverse position from waveguide edge (\SI{100}{nm} from surface) \textbf{(a)} to waveguide center \textbf{(b)}, from multimode waveguide \textbf{(b)} to single mode waveguide \textbf{(c)}, and moving further away (\SI{300}{nm} from surface) \textbf{(d)} from the waveguide surface. The waveguide widths and the relative positions of the electron beam are also labeled at the lower right corner of the panels.  
    }
    \label{fig:Green2}
\end{figure*}

\begin{figure}[t]
 	\centering
    \includegraphics[width=0.5\textwidth]{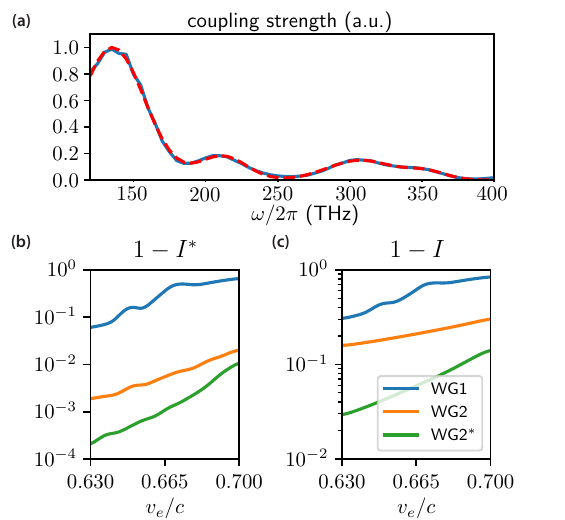}
    \caption{\textbf{(a)} An example fitting of the coupling spectrum to extract the coupling idealities. Calculated coupling ideality deviations from unity with \textbf{(b)} sideband-conditioned ($1-I^*$) and \textbf{(c)} non-conditional ($1-I$), shown with different waveguide/electron configurations (WG1:\SI{2.1}{\micro m} width; WG2:\SI{800}{nm} width; WG2*:electron beam 200nm further away from the waveguide surface), as a function of electron velocity.}
    \label{fig:ideality}
\end{figure}
Since all the physical quantities we are interested in can be related to the Green tensor of the classical Maxwell equation given the dielectric structure of interest, we numerically solved the relevant Green tensor component $G_{zz}(\bd{r},\bd{r}',\omega)$ of an infinitely long optical waveguide with finite element method (FEM). The spatial map of the imaginary part of the Green function is illustrated in Fig. \ref{fig:Green}. The Green function can be understood intuitively as the Fourier component of the optical field at frequency $\omega$ that is excited at position $\bd{r}$ by the propagating electron at position $\bd{r}'$, whereas the phase-matching condition determines whether this field constructively or destructively builds up at a given electron velocity.

The Green tensor solution of Maxwell equation is not directly supported in COMSOL, but can be retrieved by Frequency domain study with the radio frequency (RF) module. The waveguide is an air cladded \SiN slab embedded in SiO2 substrate with different geometries mentioned in the main text. Perfect matching layers at boundaries are used to prevent boundary reflections and in turn allow us to simulate an infinitely long waveguide. In order to solve for the Green function $G(\bd{r},\bd{r}',\omega)$, a point oscillating electric current dipole $\bd{J}(\omega) = \bd{p}(\omega)\delta(\bd{r}-\bd{r}_0)$ is placed above the waveguide surface at position $\bd{r}_0$ (typically \SI{100}{nm} or \SI{300}{nm}). COMSOL solves for the electric field which relates to the Green tensor as
\begin{equation*}
	\bd{E}(\bd{r}, \omega) = -4\pi i\omega G(\bd{r},\bd{r}_0,\omega)\cdot\bd{p}(\omega)
\end{equation*}
and thus if one wishes to retrieve $G_{zz}$ component one needs to orient the electric dipole $\bd{p} = p\hat{\bd{z}}$ along the z direction $\hat{\bd{z}}$, and look at the electric field z component $E_z$, such that
\begin{equation*}
	G_{zz}(\bd{r},\bd{r}',\omega) = \frac{E_z(\bd{r}, \omega)}{-4\pi i\omega p(\bd{r}',\omega)}
\end{equation*}
The results are illustrated in Fig.\ref{fig:Green}. The imaginary part of the Green function can be thought of as the spatial pattern of electron emission in the waveguide (or surrounding substrates) before the application of phase-matching condition. Given the electron velocity, the application of phase-matching
\begin{gather*}
    \bd{U}_{\hat a_\omega}(\bd{r})\propto\int dz e^{i\omega z/v_e}\mathrm{Im}[G(\bd{r};\bd{R}_0,z;\omega)\cdot\hat{ \bd z}],
\end{gather*}
retrieves the field function of the excited optical mode. The Green function along the electron trajectory is shown in Fig.\ref{fig:Green_APP}(a), where one can clearly see the bulk mode contribution near the dipole position, and spatial beatings of different waveguide modes under some conditions. 

With the optimal phase-matching condition, the coupling strength at a given optical frequency (or a discrete cavity mode) scales quadratically with the interaction length, a unique feature of guided modes co-propagating with the flying electron. For the spatial modes in the substrate bulk, the excited field is localized around the electron position. Without the benefit of constructive interference from co-propagation with the flying electron, their intensity only scales linearly with respect to interaction length. 

The total coupling strength is related to the Green function through a spatial Fourier transform, and shown in Fig.\ref{fig:Green_APP}(b), where one can identify several prominent peaks, mainly contributed from the waveguide modes, and a rising background in the Cherenkov regime ($v\gtrsim 0.7c$) of the silica substrate due to the enhanced bulk mode coupling. The Blackman window is used to eliminate the ripples from the Fourier transform due to finite simulation length, and shapes each coupling bands to a near-Gaussian shape for easier fitting of the coupling strength with a Gaussian function. The center velocity of the peaks corresponds to the optical mode phase velocity, and the bandwidth is determined by the interaction length. To improve visualization, we set the interaction length to 30 wavelengths to keep the bandwidths at different optical frequencies uniform. By sweeping the optical frequency in the simulation across the range where we have access to material permittivity, one retrieves the 2D maps shown in Fig.\ref{fig:Green2}.

With a multimode waveguide, shown in Fig.~ \ref{fig:Green2}(a-b), the effective mode index difference between the fundamental mode and higher-order modes is relatively weak at the same optical frequency, which leads to multimode electron-photon interaction within a given frequency band. When the waveguide cross-section is reduced (referred to as \emph{single-mode} waveguide), shown in Fig.\ref{fig:Green2}(c), one can enhance the mode index contrast. As a result, the mode frequency spacing is increased, such that the coupled fundamental modes are better isolated. Since most transmission electron microscopes (TEMs) have an energy resolution around \SI{0.5}{eV} (\SI{120}{THz} in optical frequency), it is important to create a large frequency spacing between the phase-matched optical modes so that the interaction with individual modes can be energy resolved.

The difference between multimode and single-mode waveguides in fiber optics is usually quantified by a V-number, a normalized frequency parameter which determines the number of modes of a step-index fiber, as
\begin{gather*}
    V = \frac{2\pi r}{\lambda}\sqrt{n_1^2-n_2^2}
\end{gather*}
where $r$ is the radial size of the core, $n_1$ is the core material index and $n_2$ is the cladding material index. For our single-mode waveguide dimension, the single-mode guiding criteria $V<2.4$ is satisfied. When used in the fiber-optic applications, such a criterion is very important for single-mode guiding. Here, our design goal is to increase the mode frequency spacing between mode families. Therefore, we only use it as a guiding principle, not as a strict criterion.

The evanescent field of the coupled higher-order modes decay faster than that of the coupled fundamental mode, as a result of the higher optical frequency. In Fig.~\ref{fig:Green2}(d), we show that one can further enhance the coupling contrast between fundamental mode and higher-order modes by placing the electron beam further away (\SI{200}{nm}) from the waveguide surface. In this way, the interaction exponentially favors the fundamental mode, at the expense of weaker interaction strength $|g_{\mathrm{TM\textsubscript{00}}}|^2$ which can be compensated for with a longer interaction length (5 times longer for the shown example).

As discussed in the main text, the Cherenkov radiation contributes as a rising background in the high velocity region. In Appendix \ref{append:substrate}, we isolate the contribution of the substrate bulk modes. 

We now quantitatively evaluate the coupling ideality to the TM\textsubscript{00} mode as a function of electron velocity. We fit the coupling spectra with Gaussian functions, illustrated in Fig.\ref{fig:ideality}(a), and calculate the conditional and non-conditional idealities as a function of electron velocities, shown in Fig.\ref{fig:ideality}(b-c).

\section{Substrate and thin film losses}\label{append:substrate}
\begin{figure*}[t]
 	\centering
    \includegraphics[width=0.75\textwidth]{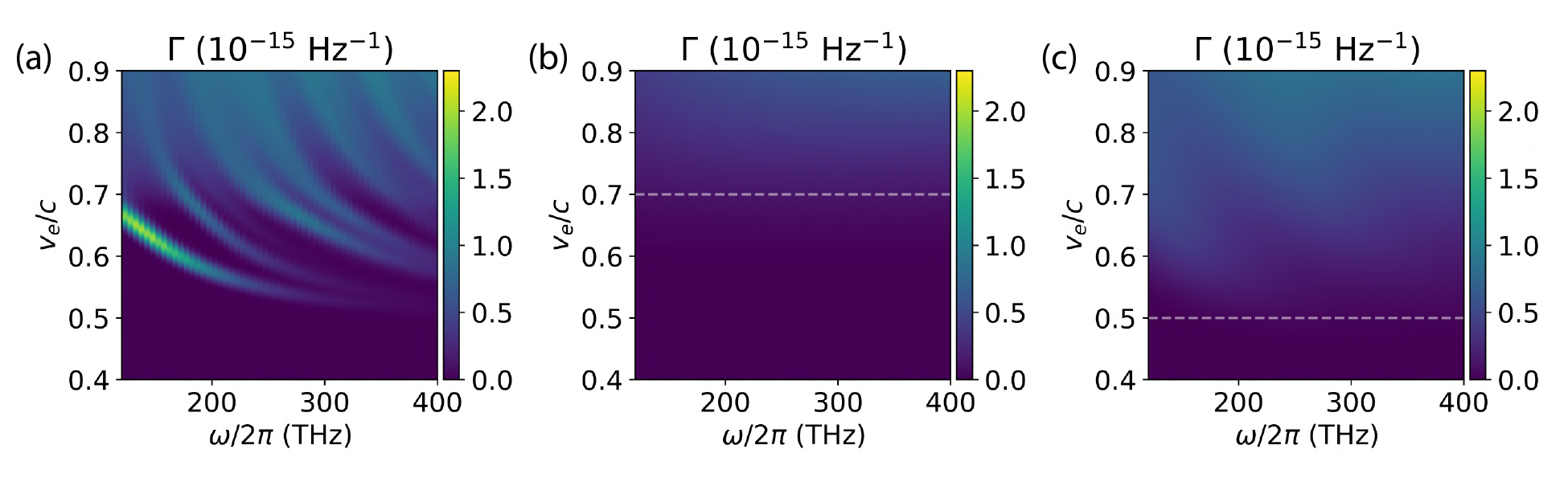}
    \caption{
    Electron energy loss spectrum for a \SI{50}{\micro\meter} interaction distance with \textbf{(a)} waveguides, \textbf{(b)} silica substrate, and \textbf{(c)} \SiN thin film on silica substrate. Notice that for a waveguide the scaling is quadratic with respect to distance and more structured, whereas for substrate and thin film the scaling is mostly linear and the emission is more broadband. The interaction with the waveguide mode will prevail over substrate and thin film over an interaction length of just a few wavelengths. The dashed gray lines are a guide to the eye showing Cherenkov regime boundaries for silica and \SiN. }
    \label{fig:2D_substrate}
\end{figure*}

Though not discussed in the main text, there are different scaling of $\Gamma(
\omega)$ for bulk substrate ($\propto L$), thin film ($\propto L\log L$), and guided modes ($\propto L^2$). We show their coupling spectrum characteristics in Fig.\ref{fig:2D_substrate} with an electron \SI{100}{nm} above the dielectric surface. It is shown that for a given frequency component $\omega$, the quadratic scaling of a guided mode will dominate the interaction. However, for a waveguide structure with linear dispersion (e.g. the one shown in Fig.\ref{fig:2D_substrate}(a)), the phase-matching condition will enforce a linear scaling of the total deposited quanta into one particular waveguide mode. But due to relatively weak dispersion of the waveguide modes, the coupling contribution from the waveguide modes dominates over substrate losses, where the latter accounts for far less than 1\% of the total coupling strength over a \SI{0.6}{eV} band with electron velocity $v_e/c\leq 0.6$. For unpatterned thin films, the total photon emission is 70\% lower than for a waveguide, and the emission is less structured and hard to collect. Note that due to the presence of chromatic dispersion, the total coupling strengths of different spatial-temporal optical modes are linearly dependent on distance. Therefore, the ratio of different coupling contributions is distance independent, and only depends on waveguide dispersion and routing, and e-beam positioning.

\section{Interaction with optical resonators}\label{append:res}
We only discussed results for open ended waveguides so far. However, there have been experiments \cite{Henke2021,feist22} that uses optical resonators with a discrete set of well-defined frequency modes instead of a continuum of frequency modes in the case of a waveguide. These well-defined modes in state of the art resonators typically have optical linewidths of tens of \SI{}{MHz}~\cite{Liu2021}, and it is therefore difficult to resolve their Green functions by sweeping the optical frequencies in FEM simulations. Nonetheless, their Green functions can be easily related to the one of open ended optical waveguides by their optical susceptibility function $\chi(\omega) = \frac{2}{\pi}\frac{F}{1+4(\omega-\omega_0)^2/\kappa^2}$ enforced by the resonator periodic boundary conditions, describing an optical resonance with center frequency $\omega$ and Finesse $F=\frac{\Delta \nu_{\mathrm{FSR}}}{\kappa}$, where the cavity free-spectral-range (FSR) is used. One can retrieve the resonator Green function $G(\omega)$ by separating the open waveguide Green function into contributions from different cavity modes (with mode field function $\bd{U}_m(\bd{r})$, details see Appendix \ref{append:MD}) 
\begin{gather*}
    G_m(\bd{r},\bd{r}',\omega)=\bd{U}_m(\bd{r'})\int d^3\bd{r}''\epsilon(\bd{r}'',\omega)\bd{U}_m^*(\bd{r''})G(\bd{r},\bd{r}'',\omega),
\end{gather*}
and multiplying the resonance susceptibilities $G(\omega)=\sum_mG_m(\omega)\chi_m(\omega)$. For a closed loop resonator structure, the resulting interaction strength $\Gamma(\omega)$ will have a narrow-linewidth comb-like structure~\cite{feist22} instead in frequency space, compared to the continuum case of an open ended waveguide, with the peak intensity enhanced by a factor of $\frac{2F}{\pi}$. 

\begin{figure}[t]
 	\centering
    \includegraphics[width=0.49\textwidth]{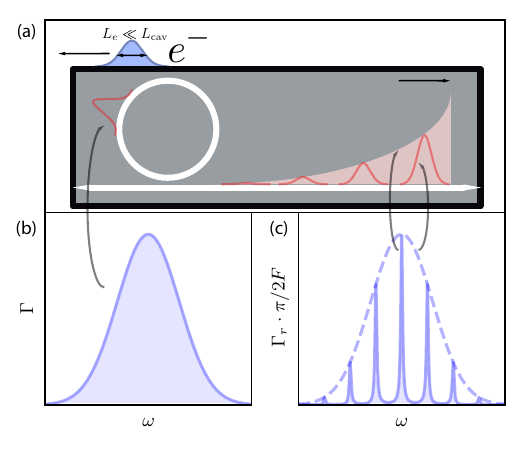}
    \caption{Comparison between free electron coupling to waveguides vs. resonators. \textbf{(a)} Illustration of electron-photon interaction mediated by an optical cavity. An optical pulse is first generated inside the cavity. Then the pulse circulates inside the cavity and couples out as a pulse train with repetition rate of FSR and an exponential decaying envelope. \textbf{(b)} Coupling spectrum for the pulse inside the resonator, indistinguishable to the one generated inside a waveguide with the same geometry. \textbf{(c)} Coupling spectrum for the out-coupled pulse trains, now exhibits a comb-like feature with frequency spacing matching the resonator FSR, peak width matching the pulse train envelope, and comb envelope matching the spectral components of each individual pulse. }
    \label{fig:SI_resonator}
\end{figure}

A comparison between a waveguide and a resonator coupling to free electrons ( $\Gamma$ and $\Gamma_r$ respectively) is illustrated in Fig.~\ref{fig:SI_resonator}. The comb-like structure in the electron energy loss spectrum results from the spectral property of the resonator that is non-local with respect to the interaction region, and is only accessible since the interaction is analyzed in the electron energy basis, whose state is also non-local in nature. However, in order to access these comb-like features in an EELS experiment, the electron-cavity characteristic interaction time (determined by the measured electron ZLP) has to be longer than the round trip time of the resonator, thus satisfying the energy-time uncertainty principle. Nonetheless, the comb-like structure can always be accessed from the optical side with a measurement time longer than the round trip time, as was shown in~\cite{feist22}.  

There is also no difference in the total coupling strength in a given mode family for the open waveguide case and the resonator case, as long as the the phase-matching bandwidth $\Delta\nu$ is much larger than $\Delta\nu_{\mathrm{FSR}}$.
The total interaction strength $|g_m|^2$ of a mode family will be considerably altered by the resonator structure when the phase-matching bandwidth $\Delta\nu$ is on the frequency scale of one FSR.
The minimum number of modes inside the phase-matching bandwidth can be estimated with $N\sim\frac{1}{|n_g-n_{\mathrm{eff}}|}$ (for common dielectric materials $\sim5-20$), so in order to access the regime where the phase-matching bandwidth is smaller than the FSR, one requires $|n_g-n_{\mathrm{eff}}|>1$, which is generally very difficult to achieve with structures using only dielectric materials. However, with common dielectric structures and careful mode dispersion engineering, the regime $N=\mathcal{O}(1)$ where the resonance structure has a small impact is accessible.

The motivation of using a resonator instead of an open waveguide is that the optical resonance frequencies are more passively stable, and the wavepackets generated from each resonator mode are generally much longer than the optical pulse length enforced by the phase-matching bandwidth from an open waveguide, and have energy density enhanced by the cavity finesse at resonant frequencies. Therefore, resonators have advantages in experiments where optical excitation needs to interfere with a reference local oscillator, and good mode-matching is required. Resonators also provide advantages in experiments where optical frequency filtering is required, since the optical density of the excitation is concentrated in frequency. We show a frequency conversion example in Appendix~\ref{append:conversion} to exploit this advantage of optical resonators to convert \SI{}{THz} broad optical excitation to \SI{}{MHz} broad optical or microwave excitation, useful to bridge interactions with superconducting qubits, mechanical oscillators and long-life-time optical qubits.

\section{Purity and Fidelity of heralded optical quantum states}\label{appendix:optical}
In this section, we derive the heralded optical states by measuring electron energy, and calculate the state fidelity and purity. 
The optical state generated when there is no sideband overlap (not a fundamental limit), and when conditioning on a narrow bandwidth around an energy slice $E_c\sim N\hbar\omega$ is
\begin{equation*}
	|\psi_{ph}\rangle = \frac{\int  \prod_{k=1}^N d\omega_k g_{\omega_k}^* \psi(E+\hbar(\sum_k\omega_k))\hat a_{\omega_k}^\dag|0\rangle}{\sqrt{\int\prod_{k=1}^Nd\omega_k |g_{\omega_k}|^2|\psi(E+\hbar(\sum_k\omega_k))|^2(N!)^2 }}
\end{equation*}
As one can see, the coefficient is a product between the electron wavefunction and the coupling coefficient. This reflects on the fact that the electron energy distribution is correlated with the frequency at which the photon is created. One will see later that this is not the case for conditional electron states, since in the no-recoil limit the frequency of the photon does not depend on the energy of the electron. 

In the first limit when electron ZLP is much narrower than phase-matching bandwidth, we can simplify the expression to
\begin{equation*}
	|\psi_{ph}\rangle = \frac{\int  \prod_{k=1}^N d\omega_k  \psi(E+\hbar(\sum_k\omega_k))\hat a_{\omega_k}^\dag|0\rangle}{\sqrt{\int\prod_{k=1}^Nd\omega_k |\psi(E+\hbar(\sum_k\omega_k))|^2(N!)^2 }}.
\end{equation*}
for single-photon states. The frequency components of the generated state are directly linked to $\phi(\omega)\propto \psi(E_c+\hbar\omega)$. Ignoring the waveguide dispersion during propagation, we have the optical waveform
\begin{equation*}
	\phi(T = t-z_\|/v) = \widetilde{\psi}(T) e^{i\omega_c T}
\end{equation*}
where it has a center frequency determined by the conditional electron energy, and an envelope profile that is exactly the time domain electron spatial profile $\widetilde{\psi}(T)$. Therefore, by shaping electron wavefunctions and conditioning on the selected sideband energy, one can transfer the electron spatial wavefunction to the optical waveform of the photonic state at a desired optical frequency. For readers familiar with optical spontaneous parametric down conversion, a similar technique is also used in heralded single photon sources ~\cite{Lvovsky01} to imprint the waveform of the pump field onto the signal field. 

For conditional multi-photon optical states, as one can imediately see from the expression, since the electron wavefunction generally can not be factorized to $N$ components $\psi(E+\hbar(\sum_k\omega_k)\neq\prod_k F(E,\omega_k)$, the conditional state can not be addressed into a Fock state of a well defined spatial-temporal mode, but since high phase-matching bandwidth usually comes with low $g$, we restrict ourselves to single-photon states in this limit. 

We defined the fidelity of the heralded single-photon state as
\begin{widetext}

\begin{equation*}
	\mathcal{F} = \left|\langle\psi_{ph}'|\psi_{ph}\rangle\right|^2 = \frac{\left|\int   d\omega  g_\omega|\psi(E+\hbar\omega)|^2\right|^2}{\left(\int d\omega |g_\omega|^2 |\psi(E+\hbar\omega)|^2\right)\left(\int d\omega |\psi(E+\hbar\omega)|^2\right) }.
\end{equation*}

\end{widetext}
For the case of long propagation, we usually end up with very narrow phase-matching bandwidth and high interaction $g$. In this case, when conditioning on the Nth energy sideband, we can simplify the expression to
\begin{equation*}
	|\psi_{ph}\rangle = \frac{\int  \prod_{k=1}^N d\omega_k g_{\omega_k}^* \hat a_{\omega_k}^\dag|0\rangle}{\sqrt{\int\prod_{k=1}^Nd\omega_k |g_{\omega_k}|^2(N!)^2 }}
\end{equation*}
where it is a well defined $N$-photon Fock state with mode profile $\phi(t)$ (see Appendix~\ref{append:state}), which is determined by both the waveguide routing and the material dispersion. To this end, one can adapt the electron positioning and velocity to shape the optical waveform.

We defined the fidelity to this state as
\begin{widetext}
\begin{equation*}
	\mathcal{F} = \left|\langle\psi_{ph}'|\psi_{ph}\rangle\right|^2 =  \frac{\left|\int  \prod_{k=1}^N d\omega_k |g_{\omega_k}|^2 \psi(E+\hbar(\sum_k\omega_k))\right|^2}{\left(\int\prod_{k=1}^Nd\omega_k |g_{\omega_k}|^2\right)\left(\int\prod_{k=1}^Nd\omega_k |g_{\omega_k}|^2|\psi(E+\hbar(\sum_k\omega_k))|^2\right) }
\end{equation*}
\end{widetext}
With more electron operations stages, we can select the electron measurement basis. After the pair-state generation, if we pass the electron through e.g. a PINEM interaction stage characterized by the scattering matrix $\mathcal{S}(\alpha)$, by conditioning on the energy sideband $|E\rangle$, we are effectively measuring under the basis $\mathcal{S}^\dag(\alpha)|E\rangle = \sum_i c_i^*|E_i\rangle$. Formally, we write down the conditional optical state as
\begin{gather*}
    |\psi_{ph}\rangle = \exp(-|g|^2/2)\sum_N\frac{c_Ng^n}{\sqrt{N!}}|N\rangle\\\hat\rho_{ph}=\langle E|\mathcal{S}(\alpha)\hat\rho\mathcal{S}^\dag(\alpha)|E\rangle
\end{gather*}
where $|g|^2=\int d\omega |g_\omega|^2$. This effectively projects the optical state into a more general state other than Fock states if one directly measures the electron energy after the pair-state preparation. For these general states, the corresponding heralded state fidelity is an average of all the involved Fock states with a correct weight
\begin{widetext}
\begin{equation*}
	\mathcal{F} = \left|\langle\psi_{ph}'|\psi_{ph}\rangle\right|^2 = \left|e^{-|g|^2}\sum_N \frac{|c_N|^2|g|^{2n}}{N!} \frac{\int  \prod_{k=1}^N d\omega_k |g_{\omega_k}|^2 \psi(E_N+\hbar(\sum_k\omega_k))}{\sqrt{\left(\int\prod_{k=1}^Nd\omega_k |g_{\omega_k}|^2\right)\left(\int\prod_{k=1}^Nd\omega_k |g_{\omega_k}|^2|\psi(E_N+\hbar(\sum_k\omega_k))|^2\right) }}\right|^2
\end{equation*}
\end{widetext}
and the same kind of weighted averaging needs to be applied to the heralded state purity as well.

The single-photon state purity can be degraded by two main effects. First, spectral overlap between different sideband orders. Second, relative bandwidth ratio between ZLP and phase-matching bandwidth. For higher-order Fock states, it's further affected by the spectral distribution of the other optical mode families as well, e.g. the electron might not be able to distinguish between a two-photon transition of the fundamental optical mode, and a single-photon transition of a higher order mode. Here, we categorize this case into the spectral overlap between electron sidebands. \\

First, let us investigate the purity degradation of the conditional photonic Fock state due to the sideband spectral overlap. In the limit of narrow phase-matching bandwidth, the optical density matrix after detection at electron energy band $\Delta E$ electron energy event, is
\begin{widetext}
\begin{equation*}
\begin{split}
	\hat\rho_{ph}  = \frac{1}{\sum_n\frac{(\int d\omega|g_\omega|^2)^n}{n!}\int_{\Delta E}dE |\psi(E+n\hbar\omega)|^2}\int_{\Delta E}dE&\left(\sum_n\frac{(-\int d\omega g_\omega^*\hat a_\omega^\dag)^n}{n!}\prod_\omega|0_\omega\rangle\psi(E+n\hbar\omega)\right)\\&\times\left(\sum_{n}\psi^*(E+n\hbar\omega)\prod_\omega\langle 0_\omega|\frac{(-\int d\omega g_\omega \hat a_\omega)^{n}}{n!}\right)
	\end{split}
\end{equation*}
The purity of this state is 
\begin{equation*}
\begin{split}
	\mathrm{Tr}\left[\hat\rho_{ph}^2\right] = \frac{\sum_{n,n'}\frac{(\int d\omega|g_\omega|^2)^{n+n'}}{n!n'!}\iint_{\Delta E}dEdE'\psi(E+n\hbar\omega)\psi^*(E+n'\hbar\omega)\psi^*(E'+n\hbar\omega)\psi(E'+n'\hbar\omega)  }{\left(\sum_n\frac{(\int d\omega|g_\omega|^2)^n}{n!}\int_{\Delta E}dE |\psi(E+n\hbar\omega)|^2\right)^2}
	\end{split}
\end{equation*}
\end{widetext}
Then we investigate the effect of finite phase-matching bandwidth, and in the limit where there is no photon sideband overlaps, the conditional single-photon Fock state is 
\begin{widetext}
\begin{equation*}
\begin{split}
	\hat\rho_{ph} = \frac{1}{\iint_{\Delta E}dEd\omega |g_\omega|^2|\psi(E+\hbar\omega)|^2}&\int_{\Delta E}dE\left(\int  d\omega g_\omega^* \psi(E+\hbar\omega)\hat a_\omega^\dag|0\rangle\right)\\&\times\left(\int d\omega g_\omega \psi^*(E+\hbar\omega)\langle0| \hat a_\omega\right)
	\end{split}
\end{equation*}
with corresponding state purity
\begin{equation*}
\begin{split}
	\mathrm{Tr}\left[\hat\rho_{ph}^2\right] = \frac{\iint_{\Delta E}dEdE'd\omega d\omega' |g_\omega|^2|g_{\omega'}|^2\psi(E+\hbar\omega)\psi^*(E+\hbar\omega')\psi^*(E'+\hbar\omega)\psi(E'+\hbar\omega')}{\left(\iint_{\Delta E}dEd\omega |g_\omega|^2|\psi(E+\hbar\omega)|^2\right)^2}
	\end{split}
\end{equation*}
\end{widetext}
In the limit of perfect electron energy resolution $\Delta E\rightarrow 0$, $\mathrm{Tr}\left[\hat\rho_{ph}^2\right]\rightarrow 1$. However, experimentally, either the ZLP consists of a statistical uncertainty, or the conditioning window can not be set arbitrarily small, due to its effect on the heralding rate. As a result, the purity is limited by both the phase-matching bandwidth, and the relative heralding bandwidth. \\

\begin{figure}[t]
 	\centering
    \includegraphics[width=0.33\textwidth]{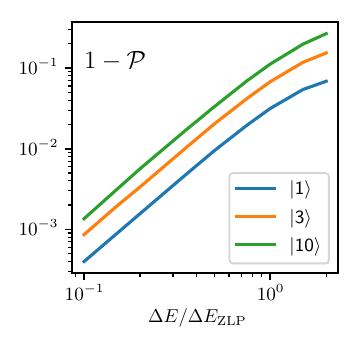}
    \caption{Heralded optical state purity vs relative heralding bandwidth at different Fock state basis. Here we assume \SI{20}{\micro\meter} interaction length. }
    \label{fig:SI_purity}
\end{figure}

\begin{widetext}

For a general $N$-photon Fock state, the density matrix of the heralded state is
\begin{equation*}
\begin{split}
	\hat\rho_{ph} = \frac{1}{\iint_{\Delta E}dE\prod_{k=1}^Nd\omega_k |g_{\omega_k}|^2|\psi(E+\hbar(\sum_k\omega_k))|^2(n!)^2}&\int_{\Delta E}dE\left(\int  \prod_{k=1}^N d\omega_k g_{\omega_k}^* \psi(E+\hbar(\sum_k\omega_k))\hat a_{\omega_k}^\dag|0\rangle\right)\\&\times\left(\int \prod_{k=1}^Nd\omega_k g_{\omega_k} \psi^*(E+\hbar(\sum_k\omega_k))\langle0| \hat a_{\omega_k}\right)
	\end{split}
\end{equation*}
with the corresponding state purity
\begin{equation*}
\begin{split}
	\mathrm{Tr}\left[\hat\rho_{ph}^2\right] = \frac{\iint_{\Delta E}dEdE'\prod_{i,j=1}^{N} d\omega_i d\omega_j' |g_{\omega_i}|^2|g_{\omega_j'}|^2\psi(E+\hbar(\sum_i\omega_i))\psi^*(E+\hbar(\sum_j\omega_j'))\psi^*(E'+\hbar(\sum_i\omega_i))\psi(E'+\hbar(\sum_j\omega_j'))}{\left(\iint_{\Delta E}dE\prod_{i=1}^Nd\omega_i |g_{\omega_i}|^2|\psi(E+\hbar(\sum_i\omega_i))|^2\right)^2}
\end{split}
\end{equation*}

\end{widetext}
To illustrate the impact of relative heralding bandwidth on state purity, we show the overall scaling of $1-\mathcal{P}\propto \gamma^2$ in the limit of small heralding bandwidth in Fig.~\ref{fig:SI_purity}. For a general state $|\psi_{ph}\rangle = \sum c_N |N\rangle$, which consists of a coherent superposition of different Fock states $|N\rangle$, as discussed before in the state fidelity calculation, the purity is a $|c_N|^2$ weighted average of each individual Fock state component, shown as
\begin{widetext}
\begin{gather*}
    \mathrm{Tr}\left[\hat\rho_{ph}^2\right] = \sum_N|c_N|^4\mathcal{P}_N+\sum_{N,N'}|c_N|^2|c_N'|^2\mathcal{P}_{N,N'}\\
	\mathcal{P}_{N,N'} = \left(\iint_{\Delta E}dEdE'\prod_{i=1}^{N}\prod_{j=1}^{N'} d\omega_i d\omega_j' |g_{\omega_i}|^2|g_{\omega_j'}|^2\right.\\
	\times\left.\psi(E+\hbar(\sum_i\omega_i-N\overline{\omega}))\psi^*(E+\hbar(\sum_j\omega_j'-N'\overline{\omega}))\psi^*(E'+\hbar(\sum_i\omega_i-N\overline{\omega}))\psi(E'+\hbar(\sum_j\omega_j'-N'\overline{\omega}))\right)\\
	\times\left(\iint_{\Delta E}dE\prod_{i=1}^Nd\omega_i |g_{\omega_i}|^2|\psi(E+\hbar(\sum_i\omega_i-N\overline{\omega}))|^2\right)^{-1}\left(\iint_{\Delta E}dE\prod_{j=1}^{N'}d\omega_j' |g_{\omega_j'}|^2|\psi(E+\hbar(\sum_j\omega_j'-N'\overline{\omega}))|^2\right)^{-1},
\end{gather*}
\end{widetext}
where $\mathcal{P}_N$ is the purity of the $N$th Fock state component, $\mathcal{P}_{N,N'}$ is the purity of the off-diagonal terms, and $\overline{\omega}$ is the center frequency of the optical state.

\section{Purity of heralded electron state}
\label{appendix:electron}
In this section, we consider heralding schemes that generate complex electron states, and derive the expression of purity of the heralded electron state by the optical measurement. 
With multiple single-photon detectors, if one conditions on a $N$-photon counting event, one projects the electron into a corresponding energy state that loses a equal amount of energy. However, these types of photon counting measurements can not project the electron into a coherent superposition of multiple sideband states. This is the result of the chosen measurement operator $a^\dag a$, whose eigenstates are Fock states. However, one can select a measurement basis to project the electron onto a more general state. These operations require high detection efficiencies (no information loss), which has to be considered carefully when applying optical elements e.g. spectral filters. The first simple method to change the measurement basis would be to combine the signal with a strong local oscillator through a high aspect ratio beam splitter. This modifies the detection from photon-number detection to field detection in the basis of displaced Fock states. With the ability to mode-match to an optical reference field, homodyne types of detection can also be realized. In the setting where the signal field is split and detected by two homodyne in orthogonal quadratures, the measurement is effectively under the coherent state basis.
More sophisticated operation can be done with an atomic system to provide arbitrary measurement basis.

In the special case of measuring in coherent state basis $|\alpha\rangle$, we can derive the effective modulation applied on the electron wavefunction as
\begin{gather*}
    A(z) = \sum_N\frac{(ge^{i\frac{\omega}{v}z})^N}{\sqrt{N!}}\frac{\alpha^N}{\sqrt{N!}}
    = e^{g\alpha e^{i\frac{\omega}{v}z}}
\end{gather*}
which is effectively a direct density modulation of
\begin{gather*}
    |A(z)|^2 = e^{2|g\alpha|\cos(\frac{\omega}{v}z+\theta_{g\alpha})}
\end{gather*}
Now, we consider a general heralded electron state, with the $N$th sideband density matrix component as
\begin{widetext}
\begin{equation*}
\begin{split}
	\hat\rho_e = \frac{1}{\iint_{\Delta \omega}dE\prod_{k=1}^Nd\omega_k |g_{\omega_k}|^2|\psi(E+\hbar(\sum_k\omega_k))|^2}&\int_{\Delta \omega}\prod_{k=1}^N d\omega_k|g_{\omega_k}|^2\left(\int dE \psi(E)|E-\hbar(\sum_k\omega_k)\rangle \right)\\&\times\left(\int dE\psi^*(E)\langle E-\hbar(\sum_k\omega_k)|\right)
	\end{split}
\end{equation*}
\end{widetext}

Note that for each sideband component, the corresponding electron wavefunction is not shaped by the optical detection and maintains the original shape. This is in sharp contrast to heralding optical state by measuring electron energies. The difference is that the electron energy loss heavily depend on optical frequency, but the optical frequency does not depend on electron energy under the no-recoil approximation. Therefore, any measurement on the frequency of the created photons simply translates the original electron energy state down by a corresponding photon energy. Because of that, we do not define the fidelity of the electron wavefunction in the limit of perfect photon frequency resolution.\\
\begin{widetext}

We proceed to calculate the state purity of the $N$th electron sideband component as
\begin{equation*}
\begin{split}
	\mathrm{Tr}\left[\hat\rho_e^2\right] = \frac{\iint_{\Delta \omega}dEdE'\prod_{i,j=1}^{N} d\omega_i d\omega_j' |g_{\omega_i}|^2|g_{\omega_j'}|^2\psi(E+\hbar(\sum_i\omega_i))\psi^*(E+\hbar(\sum_j\omega_j'))\psi^*(E'+\hbar(\sum_i\omega_i))\psi(E'+\hbar(\sum_j\omega_j'))}{\left(\iint_{\Delta \omega}dE\prod_{i=1}^Nd\omega_i |g_{\omega_i}|^2|\psi(E+\hbar(\sum_i\omega_i))|^2\right)^2}
\end{split}
\end{equation*}

\end{widetext}
where the same weighted average needs to be applied for a general state with weights $|c_N|^2$, similar to the case of heralded optical states.

\section{Optical waveform generated from the electron-photon interaction}\label{append:state}
In this section, we derive the waveform of the conditional optical state when considering the waveguide dispersion. 

The composite quantum state after the electron-photon interaction is
\begin{widetext}

\begin{gather*}
    |\psi\rangle = \exp\left(\int d\omega g_\omega \hat b_\omega^\dag \hat a_\omega - h.c. \right)|\psi_e\rangle|0\rangle
    =e^{-\frac{\int d\omega|g_\omega|^2}{2}}e^{-\int d\omega g_\omega^* \hat b_\omega \hat a_\omega^\dag}e^{\int d\omega g_\omega \hat b_\omega^\dag \hat a_\omega}|\psi_e\rangle|0\rangle\\
    = e^{-\frac{\int d\omega|g_\omega|^2}{2}}e^{-\int d\omega g_\omega^* \hat b_\omega \hat a_\omega^\dag}|\psi_e\rangle|0\rangle
    = e^{-\frac{\int d\omega|g_\omega|^2}{2}}\sum_n\frac{\left(-\int d\omega g_\omega^* \hat b_\omega \hat a_\omega^\dag\right)^n}{n!}|\psi_e\rangle|0\rangle
\end{gather*}

\end{widetext}
when conditioned on the $n$th energy sideband of the electron state (with electron ZLP much wider than the coupling bandwidth to the optical modes), the heralded optical state is
\begin{gather*}
    |\psi_{ph}\rangle \sim \left(-\int d\omega g_\omega^* \hat a_\omega^\dag\right)^n|0\rangle.
\end{gather*}
When the interaction is dominated by the coupling to a single optical mode family, one can generate Fock state of a well-defined spatial-temporal mode as
\begin{gather*}
    |\psi_{ph}\rangle \sim \left(\hat a_m^\dag\right)^n|0\rangle\\
    \hat a_m = \int_{\Delta\omega_m} d\omega \phi_m(\omega)\hat a_\omega\\
    \phi_m(\omega) = \frac{g_{\omega,m}^*}{g_m^*}\\
\end{gather*}
From these results, one can derive the temporal field profile function of this spatial-temporal mode, as it may concern experiments that require waveform shaping. Straight from the definition, one gets
\begin{gather*}
\phi_m(\bd{r},t) \propto \iint dz d\omega e^{i\omega(z/v_e-t)}\tilde{U}_{m,z}^*(\bd{R}_0,z,\omega)\tilde{\bd{U}}_{m}(\bd{r},\omega).
\end{gather*}
When chromatic dispersion of the frequency modes is ignored, one retrieves the waveform shown in the main text. When dispersion is considered, one can further remove the frequency dependence of the mode profile functions by assuming an open waveguide (e.g. no sharp frequency response in the phase-matched region) and up to second order dispersion $\beta$, 
\begin{gather*}
\tilde{\bd{U}}_{m}(\bd{r},\omega)\approx \tilde{\bd{U}}_{m}(\bd{r},\omega_m)e^{i(\omega-\omega_m)r_{\|}/v_g}e^{i\beta(\omega-\omega_m)^2r_{\|}/v_g},
\end{gather*}
where $\omega_m$ is the center frequency of the pulse, selected so that the phase velocity at $\omega_m$ matches the electron velocity $v$, $v_g\lesssim v$ is the corresponding group velocity, and $r_{\|}$ is the longitudinal coordinate along the waveguide trajectory. One can then rewrite the expression as 
\begin{widetext}

\begin{gather*}
\phi_m(\bd{r},t) \propto \iint dzd\omega e^{i\frac{(\omega-\omega_m)}{v}(z-\frac{v}{v_g}(R_\|(z)-r_\|)-vt)}
e^{i\beta\frac{(\omega-\omega_m)^2}{v_g}(r_\|-R_\|(z))}e^{i\omega_m(z/v_e-t)}\tilde{U}_{m,z}^*(\bd{R}_0,z,\omega_m)\tilde{\bd{U}}_{m}(\bd{r},\omega_m)\\
\propto \int dz e^{i\frac{\left(z-\frac{v}{v_g}(R_\|(z)-\tilde{r}_\|)\right)^2}{4\beta (r_\|-R_\|(z)) v^2/v_g}}\frac{e^{i\frac{\pi}{4}\mathrm{sgn}(\beta (r_\|-R_\|(z)))}}{\sqrt{|\beta (r_\|-R_\|(z))|}}
e^{i\omega_m(z/v_e-t)}\tilde{U}_{m,z}^*(\bd{R}_0,z,\omega_m)\tilde{\bd{U}}_{m}(\bd{r},\omega_m)\\
\propto \int dz K(z,\bd{r},t)\overline{\tilde{U}}_{m,z}^*(\bd{R}_0,z,\omega_m)\tilde{\bd{U}}_{m}(\bd{r},\omega_m)e^{-i\omega_mt}
\end{gather*}

\end{widetext}
where $\tilde{r}_\|\equiv r_\|-v_g t$ is the waveform coordinate in the optical pulse frame with group velocity $v_g$. $\overline{\tilde{U}}_{m,z}^*(\bd{R}_0,z,\omega_m)$ is the mode envelope profile at wave vector $\omega_m/v$. The integral kernel
\begin{gather*}
K(z,\bd{r},t) \equiv  e^{i\frac{\left(z-\frac{v}{v_g}(R_\|(z)-\tilde{r}_\|)\right)^2}{4\beta (r_\|-R_\|(z)) v^2/v_g}}\frac{e^{i\frac{\pi}{4}\mathrm{sgn}(\beta (r_\|-R_\|(z)))}}{\sqrt{|\beta (r_\|-R_\|(z))|}}
\end{gather*}
represents a phase scrambling around the waveform coordinate $\tilde{r}_\|$ with a bandwidth of $\sim|\beta(r_\|)v^2/v_g|$, due to the presence of second order dispersion. 
One can get physical intuition of the waveform in the limit of weak dispersion ($\beta\rightarrow 0$), where one can approximate the integral kernel with a Dirac delta function,
\begin{widetext}

\begin{gather*}
\phi_m(\bd{r},t)
\propto \int dz \delta(z-\frac{v}{v_g}(R_\|(z)-\tilde{r}_\|))
\overline{\tilde{U}}_{m,z}^*(\bd{R}_0,z,\omega_m)\tilde{\bd{U}}_{m}(\bd{r},\omega_m)e^{-i\omega_mt}
\propto \sum_i\frac{\overline{\tilde{U}}_{m,z}^*(\bd{R}_0,z_i,\omega_m)}{|R_{\|\partial z}(z_i)-\frac{v_g}{v}|}\tilde{\bd{U}}_{m}(\bd{r},\omega_m)e^{-i\omega_mt}
\end{gather*}

\end{widetext}
where $z_i(\bd{r},t):\quad \frac{z_i}{v}-\frac{1}{v_g}(R_\|(z_i)-\tilde{r}_\|)=0$ are the spatial $z$ coordinates where the vacuum field contributes the most through the phase-matching condition to the generated field at $\bd{r}$ coordinate at time $t$. Therefore, the excited optical profile in the time domain is easily connected to the envelope of the optical mode field profile $\tilde{U}_{m,z}(\bd{R}_0,z,\omega_m)$ along the electron propagation direction, when the mode dispersion is sufficiently weak. In the exact limit $\beta = 0$, there can be unphysical scenarios when $|R_{\|\partial z}(z_i)-\frac{v_g}{v}|=0$, which corresponds to the infinite phase-matching bandwidth. However, in physical materials, the phase-matching bandwidth is always finite.

The mode dispersion during pulse propagation will cause pulse shortening or broadening by shifting the phase of different frequency components and leaving the amplitude unchanged. This can be easily corrected and is not a fundamental limit to construct an arbitrary waveform. Therefore, one can structure any desired optical waveform $\phi(\bd{r},t)$ by positioning the electron beam trajectory on an optical waveguide with a tailor-made waveguide structure.

\section{Imprinting electron wavefunction onto the optical waveform in an interferometric fashion}\label{appendix:interferometry}
In the regime where the electron energy spread is much narrower than the phase-matching bandwidth, the spatial-temporal mode is defined completely by the electron wavefunction, see Appendix~\ref{appendix:optical}, with a phase contribution from the coupling coefficient $g_{\omega_c=-E_c/(N\hbar)}$, where $N$ is the sideband order. Whenever an electron interacts with an optical mode, and measured on the $N$th electron energy sideband at the energy $E_c$, it's equivalent to apply an operator
$\hat S \propto \psi(E_c+\hbar\omega)\left(e^{i\theta_{g}}\hat a_{\omega}^\dag\right)^N$ onto the optical state. When the electron interacts with two waveguides in sequential manner (identical geometry), the operator is 
\begin{gather*}
    \hat S \propto \int d\omega\psi(E+\hbar\omega)\left(e^{i\theta_{g_1}}\hat a_{1,\omega}^\dag+e^{i\theta_{g_2}}\hat a_{2,\omega}^\dag\right)^N
\end{gather*}
where $a_i$ is the spatial-temporal mode on waveguide $i=1,2$, and the phase is determined by the the reference point from both the optical side and the electron side. When the two interaction regimes (positions of waveguides) are separated by a spatial distance that correspond to an electron propagation time $\Delta t$, we have the following phase relation of the coupling coefficients
\begin{gather*}
    e^{-i\hat H_0\Delta t}\hat S_{\mathrm{e\text{-}ph}}(g_\omega) e^{-i\hat H_0\Delta t}=\hat S_{\mathrm{e\text{-}ph}}(g_\omega e^{i\omega\Delta t}).
\end{gather*}
where $\hat H_0$ is the electron free evolution Hamiltonian.
If we assume the optical phase references are set to $0$, we can rewrite the scattering as
\begin{gather*}
    \hat S \propto \int d\omega\psi(E+\hbar\omega)\left(\hat a_{1,\omega}^\dag+e^{i\omega\Delta t}\hat a_{2,\omega}^\dag\right)^N
\end{gather*}
As one can see, this interaction projects the optical state into a quantum coherent spatial superposition state of two spatially separated waveguides. In the case that the two waveguide are connected to a 50:50 beam splitter to form an interferometer, the photon flux difference $f(t)$ can be expressed as
\begin{gather*}
    f(t) \propto \mathrm{Re}\left[\widetilde{\psi}(t)\widetilde{\psi}^*(t+\Delta t)e^{i\omega_c\Delta t}\right],
\end{gather*}
which forms an effective optical interferometer of the time delay the electron experiences between two interaction stages, but can also be induced by an external potential. Here, to extract the delay time, one can simply look at the photon counting record at different electron energy records $E_c$ which determines $\omega_c$. Therefore, compared to conventional optical interferometry where one has to scan the laser frequencny over a very broad range to resolve length difference to the order of a few wavelengths, here, we exploit the very wide electron emission bandwidth to get an accurate length difference, just by looking at different electron energy records. Note that here the imprinted electron wavefunction only provides a profile function with an effective optical delay (convenient for automatic mode matching between the two arms), with its original fast evolving phase unobservable. Therefore, for a phase object positioned between the two waveguides, the interferometer sensitivity is on the optical wavelength scale, not on the scale of the de Broglie wavelength of the electrons.

\section{Frequency conversion using resonator structures}\label{append:conversion}
As discussed in the previous section, optical resonators provide unique advantages over straight waveguides in terms of the concentrated optical density. Here we show an example scheme to use on-chip ring resonator structures to convert the \SI{}{THz}-broad optical excitation from the electron-photon interaction to a \SI{}{MHz}-broad optical/electrical excitation, limited by the optical resonator linewidth. Here we define the spatial-temporal mode for the optical excitation of a resonator as $\hat a^\dag = \int d\omega\sum_i\phi_i(\omega)\hat a_\omega^\dag$, where each $\phi_i(\omega)$ is a Lorentzian centered around $\omega_i+\omega_m$ with $\omega_i$ the pump center frequency. We also define the microwave excitation as $\hat c^\dag \propto \int d\omega \phi(\omega)\hat c_\omega^\dag$, centered around $\omega_m$. In the ideal case that all the optical azimuthal modes are identical in their frequency components, we have $\phi_i(\omega+\omega_i) = \phi(\omega)$.

Near-unity efficiency optical to microwave conversion was demonstrated in $\chi_2$ type materials~\cite{Sahu2022,Fan18}. We assume a ring structure optical resonator (conversion ring) with a strong $\chi_2$ nonlinearity and a relatively high optical quality factor, with identical cavity azimuthal modes of two orthogonal polarizations (e.g. TE and TM fundamental modes) with approximately the same FSR, and a frequency spacing $\sim\omega_m$ between these two mode families that matches the microwave mode frequency $\omega_m$. Practically, the frequency matching is only required between a few optical resonances, since the electron-photon optical excitation from a ring resonator can occupy only 5-20 azimuthal modes. We use this resonator for frequency down conversion of the broadband photon excitation $\hat a^\dag$ at frequencies $\omega_i+\omega_m$ from the electron-photon interaction, with a specially structured local oscillator pump at frequencies $\omega_i$. We further assume that there are two more rings on the chip, with matched optical frequencies of the modes of interest. We use one of the rings for electron-photon interaction, where the photon excitation is generated on the TM polarization at frequencies $\omega_i+\omega_m$ (signal field). This ring should be designed with the highest quality factor possible, since its linewidth determines the microwave linewidth $\phi(\omega)$, and should be narrower than the linewidth of the conversion ring. We use the other ring for generating or filtering a structured continuous wave (CW) optical pump on the TE polarization with frequencies $\omega_i$ (LO field). We combine LO and signal field through a polarization beam splitter, and send them into the conversion ring for frequency down conversion. 

In principle, the signal field and LO field does not have to be orthogonal in polarization to enable efficient and low-noise frequency conversion. The orthogonal polarizations considered here are intended to prevent spectral leakage of the LO field to the signal mode, even though they can be sufficiently separated in frequency. 

When the conversion ring is pumped by the LO field with cooperativity $C=1$ in each pump-field mode pairs, the signal frequency component $\phi_i(\omega)$ at the azimuthal mode at frequency $\omega_i+\omega_m$ is converted to a microwave photon at frequency $\omega_m$ with frequency component $\phi(\omega)$ and conversion efficiency $\eta=100\%$. When all the azimuthal modes of the conversion ring convert their signal field components down to the same microwave frequency $\omega_m$ with unity efficiencies, the original signal pulse with \SI{}{THz}-broad frequency components from the electron-photon interaction will be converted to a single microwave mode excitation at frequency $\omega_m$ with frequency width down to \SI{}{MHz} with a unity efficiency, and at the same time generate a pump field photon with \SI{}{THz}-broad frequency component due to the energy conservation.

Here, we formally analyze this conversion process. We define the scattering matrix of the conversion process with a multi-mode three-wave mixing operator
\begin{gather*}
    \hat S_{\mathrm{TWM}} = e^{\sum_i\int \beta_i(\omega,\omega_i)\hat a_{\omega+\omega_i}\hat c_\omega^\dag \hat d_{\omega_i}^\dag-h.c.}
\end{gather*}
where $\omega_i$ is the frequency of each CW pump comb tooth, and $\beta_i$ the coupling constant amplified by the pump field. The coupling constant $\beta_i(\omega,\omega_i)$ contains the conversion frequency response of each pump-signal mode pair, including effects e.g. phase-matching, cavity responses of the signal and pump field , and microwave cavity response. The operators are the signal field annihilation $\hat a_\omega$, the microwave field creation operator $\hat c_\omega^\dag$, and the pump field operator $\hat d_{\omega_i}^\dag=\sum_{N_{\omega_i}}|N_{\omega_i}+1\rangle\langle N_{\omega_i}|$ which is specially defined high-up in the photon ladder, similar to the electron ladder operator in the no-recoil limit, that represents the addition of a photon to the classical coherent pump field $|\alpha_i\rangle$ with frequency $\omega_i$. In the single mode conversion case, $\hat d^\dag$ can be ignored. However, in the multi-mode conversion considered here, neglecting $\hat d^\dag$ leads to non-unitary operations. Under such an operation, the state of the system changes to
\begin{widetext}
\begin{gather*}
    \hat S_{\mathrm{TWM}}|\psi_{ph},\psi_m, \psi_{\mathrm{pump}}\rangle = \hat S_{\mathrm{TWM}}F(\hat a^\dag)|0_a,0_c,\alpha_i\rangle \\
    = F(\hat S_{\mathrm{TWM}}\hat a^\dag\hat S_{\mathrm{TWM}}^\dag)\hat S_{\mathrm{TWM}}|0_a,0_c,\alpha_i\rangle = F(\hat S_{\mathrm{TWM}}\hat a^\dag\hat S_{\mathrm{TWM}}^\dag)|0_a,0_c,\alpha_i\rangle\\
    \hat S_{\mathrm{TWM}}\hat a^\dag\hat S_{\mathrm{TWM}}^\dag = \sum_i\phi_i(\omega+\omega_i)\sin(|\beta_i(\omega,\omega_i)|)e^{i\theta_{i}}\hat d_{\omega_i}^\dag \hat c_\omega^\dag + \sum_i \phi_i(\omega)\cos(|\beta_i(\omega-\omega_i,\omega_i)|)\hat a_\omega^\dag
\end{gather*}
\end{widetext}
where $\hat a^\dag = \int d\omega\sum_i\phi_i(\omega)\hat a_\omega^\dag$ is used, $\theta_i = \mathrm{arg}[\beta_i(\omega,\omega_i)]$, and also $|\psi_{ph}\rangle = F(\hat a^\dag)|0_a\rangle$. In the ideal case where $\phi_i(\omega+\omega_i)=\phi(\omega)G_i$ is separated to the cavity density of states $\phi(\omega)$ and the electron-photon phase-matching coefficient $G_i$ at signal mode $i$, with unity cooperativity $\beta_i(\omega,\omega_i)=\pi/2$ at every pump-signal mode pair over the frequency components of interest, the state is simplified to
\begin{gather*}
    |\psi_m,\psi_{\mathrm{pump}}\rangle = F(\hat c^\dag \hat d^\dag)|0_c,\alpha_i\rangle
\end{gather*}
where 
\begin{gather*}
    \hat c^\dag = \int d\omega \sqrt{\sum_i|G_i|^2}\phi(\omega)\hat c_\omega^\dag\\
    \hat d^\dag = \sum_i\frac{G_i}{\sqrt{\sum_i|G_i|^2}}e^{i\theta_i}\hat b_{\omega_i}^\dag
\end{gather*}
Since the pump field is a strong coherent field and maintains a unity overlap with a photon-added state, we can trace out the pump field state space, and arrive at
\begin{gather*}
    |\psi_m\rangle = F(\hat c^\dag)|0_c\rangle
\end{gather*}
where the state of the signal field $|\psi_{ph}\rangle = F(\hat a^\dag)|0_a\rangle$ is transferred to the microwave field, with frequency components limited by the signal cavity density of states $\propto \phi(\omega)$.

Practically, the amplitude and frequency components of the LO field has to be specifically shaped, but since the optical excitation from the electron-photon interaction can occupy only $5$-$20$ optical modes with the maximum interaction length achievable with a racetrack resonator geometry, it is possible to shape a reasonably accurate LO field with a soliton~\cite{Kippenberg18} or an electro-optic comb~\cite{Zhang2019} source and a frequency shaper~\cite{Weiner2000}.   

There are currently two types of main-stream optical-to-microwave converters~\cite{QuICs}. One type uses $\chi_2$ optical nonlinearity to directly convert signals from the optical domain to the microwave domain, as is considered here. The other type~\cite{Higginbotham2018} uses a mechanical oscillator as an intermediate stage to first convert the optical signal to a mechanical signal using optomechanical couplings, and then from the mechanical signal to a microwave signal using electro-mechanical couplings. Both types of systems have shown near-unity conversion efficiency and low added noise, but the mechanical one suffers from the low conversion bandwidth typically at \SI{}{kHz} level (not possible to achieve $\beta_i(\omega,\omega_i)\sim\pi/2$ over the cavity bandwidth $\sim$\SI{}{MHz}), limited by the electro-mechanical and optomechanical coupling rates. Therefore, we only consider $\chi_2$-type optical-to-microwave converters for our scheme since they offer broadband transductions, essential to convert all the frequency components of the optical excitation to the microwave domain. For a realistic lithium niobate ring resonators with \SI{50}{\micro\meter} radius, the microwave frequency at \SI{4}{GHz}, with optical and microwave cavity linewidths at $\kappa_{ph/m}/2\pi\sim \SI{10}{MHz}$, the estimated required pump power to reach $C_i=1$ is reasonable at $P_i\sim\SI{100}{\micro\watt}$~\cite{clement16}.

Similar types of schemes can also down-convert the signal to an optical excitation with \SI{}{MHz} linewidth. But generally, a relatively uniform mode spacing over multiple azimuthal modes at optical frequency scale is harder to design for triply resonant schemes. 

\section{Beyond no-recoil approximation}\label{appendix:recoil}
In the no-recoil limit, the coupling coefficient $g_\omega$ is not energy dependent, as is the case for the scenarios we discussed in the current study. However, the recoil effect will be significant with multi-stage operations with sufficient spatial separation, where the effect of electron energy-momentum dispersion kicks in. 

Here, we consider two stages separated by distance $\Delta L$, with near-point-like interaction regions. The coupling coefficient for the first stage is
\begin{gather*}
    g_{1,\omega}\propto \int dz e^{-i\frac{\omega}{v}z}U_{1,z}(z)
\end{gather*}
where $U_{1,z}(z)$ is the optical mode function along $\hat z$ direction. At the first stage the recoil effect does not play a significant role yet. However, at the second stage that's $\Delta L$ away, the electron energy dispersion $\hbar k = \sqrt{E^2/c^2-m^2c^2}$ changes the phase-matching integral of the off-diagonal element $|E\rangle\langle E+\hbar\omega|$ to
\begin{gather*}
    g_{2,\omega}(\Delta E = E-E_0)\propto \int dz e^{-i(\frac{\omega}{v}-2\pi(\frac{2\Delta E}{\hbar\omega }+1)\frac{1}{z_T})z}U_{2,z}(z)\\
    \approx e^{i2\pi(\frac{2\Delta E}{\hbar\omega }+1)\frac{\Delta L}{z_T}}\int dz e^{-i\frac{\omega}{v}z}U_{2,z}(z)
\end{gather*}
where now the coupling coefficient accumulates an energy-dependent phase $e^{i2\pi(\frac{2\Delta E}{\hbar\omega }+1)\frac{\Delta L}{z_T}}$. Here $z_T = 4\pi m v^3 \gamma^3/\hbar\omega^2$ is the Talbot distance and $\gamma$ is the Lorentz factor. $\Delta E$ is the distance to the original center electron energy $E_0$, which is the reference point of the dispersion quadratic expansion. In the case of $v/c\sim 0.65$ and $\omega\sim  2\pi\cdot\SI{2e14}{Hz}$, $z_T\sim \SI{1}{m}$. For the largest photon number $|N=10\rangle$ discussed in the current manuscript, and a typical photonic chip length $\Delta L\sim \SI{5}{mm}$, the phase accumulation is $\theta = 0.1\cdot 2\pi$, therefore negligible in our discussion. We anticipate the effect to dominate, e.g. when two photonic chips are involved and with sufficient separation $\Delta L \sim \SI{1}{m}$. When that is indeed the case, we can still use the scattering matrix with the no-recoil approximation (used in the current manuscript) at the second stage, but add a propagation matrix
\begin{equation*}
	\hat S_{\mathrm{prop}} = \int dE\exp\left[ \frac{-i2\pi\Delta L(E-E_0)^2}{4\pi \hbar m v^3\gamma^3} \right]|E\rangle\langle E|,
\end{equation*}
to account for the recoil-induced phase accumulation of different energy components when arriving at the second stage. The phases accumulate differently between different energy components due to their different group velocities, and therefore there is an effective timing difference of their arrival at the second stage which is treated as a point interaction with no-recoil approximation. This is consistent with literature that uses Schr\"{o}dinger equations to solve for the wavefunction evolution between two interaction stages~\cite{Yalunin2021}, and can also explain effects observed in double-PINEM-type experiment~\cite{Priebe2017}. 

However, our assumption of point-like interaction will break down when a single interaction is sufficiently long e.g. $L>\SI{10}{cm}$, and with transitions involving Fock states $|N>10\rangle$. Then a single scattering matrix including the recoil effect has to be used to account for the dispersive phase accumulation during the interaction, as
\begin{gather*}
    \widetilde{\hat S}_{\mathrm{e\text{-}ph}}=\exp\left[ \int d\omega g_\omega \widetilde{\hat b}_\omega^\dag \hat a_\omega - h.c. \right]
\end{gather*}
where the modified electron energy lowering operator is
\begin{gather*}
    \widetilde{\hat b}_\omega = \int dE |E\rangle \langle E+\hbar\omega | \frac{\int dz e^{i(\frac{\omega}{v}-2\pi(\frac{2(E-E_0)}{\hbar\omega }+1)\frac{1}{z_T})z}U_{z}^*(z)}{\int dz e^{i\frac{\omega}{v}z}U_{z}^*(z)}
\end{gather*}
and the no-recoil coupling coefficient $g_\omega$ is used. For non-guided electron beam with divergence angle around \SI{0.2}{mrad}, and an electron-surface gap \SI{100}{nm}, the longest propagation on chip is restricted to about \SI{1}{mm}. To achieve long enough distance such that recoil effect kicks in, an on-chip electron guiding structure~\cite{Shiloh2021,Niedermayer2020} is required.

\bibliographystyle{apsrev4-2}
\bibliography{main}

\end{document}